\documentclass[10pt]{iopart}

\usepackage{graphicx}
\usepackage{amssymb,amsfonts,dsfont}
\usepackage{color}
\usepackage{array}
\usepackage{upgreek}
\usepackage{mathrsfs}
\usepackage[normalem]{ulem}

\expandafter\let\csname equation*\endcsname=\relax
\expandafter\let\csname endequation*\endcsname=\relax
\usepackage{amsmath}

\newcommand{\red}{\color[rgb]{1,0,0}}

\newcommand{\gens}{g_\mrm{ens}}
\newcommand{\ordre}{\mathcal O}
\newcommand{\SNR}{\mrm{SNR}}

\newcommand{\Hb}[1]{\overline H^{\raisebox{-3pt}{\scriptsize{$(#1)$}}}}

\newcommand{\Hbc}[1]{\overline H^{\protect\raisebox{-3pt}{\scriptsize{$(#1)$}}}} 
\newcommand{\Lb}[1]{\overline {\mathcal L}^{\raisebox{-3pt}{\scriptsize{$(#1)$}}}}

\newcommand{\ket}[1]{\vert #1 \rangle}
\newcommand{\bra}[1]{\langle #1 \vert}
\newcommand{\braket}[2]{\langle #1 \vert #2 \rangle}
\newcommand{\mean}[1]{\langle #1 \rangle}

\newcommand{\liouv}{\mathcal{L}}

\newcommand{\proj}[1]{\ket{#1}\bra{#1}}
\newcommand{\be}{\begin{equation}}
\newcommand{\ee}{\end{equation}}

\newcommand{\al}{\alpha}

\newcommand{\g}{\gamma}
\newcommand{\G}{\Gamma}
\newcommand{\dt}{\delta}
\newcommand{\D}{\Delta}

\newcommand{\ve}{\varepsilon}

\newcommand{\q}{\theta}

\newcommand{\vtt}{\vartheta}

\newcommand{\kp}{\kappa}

\newcommand{\Ld}{\Lambda}

\newcommand{\s}{\sigma}

\newcommand{\ch}{\chi}

\newcommand{\y}{\psi}

\newcommand{\w}{\omega}
\newcommand{\W}{\Omega}

\newcommand{\barr}[1]{\begin{array}{#1}}
\newcommand{\earr}{\end{array}}
\newcommand{\eul}[1]{\mathrm{e}^{#1}}
\newcommand{\mrm}[1]{\mathrm{#1}}
\newcommand{\eq}[1]{(\ref{#1})}
\newcommand{\hc}{\mrm H.\mrm c.}

\newcommand{\mvec}[1]{\boldsymbol{\mathrm{#1}}}	

\newcommand{\matquatre}[1]{\left( \barr{cccc} #1 \earr \right)}

\begin{document}

\title[]{Hamiltonian engineering for robust quantum state transfer and qubit readout in cavity QED}

\author{F\'elix Beaudoin$^1$, Alexandre Blais$^{2,3}$, W. A. Coish$^{1,3}$}
\address{$^1$ Department of Physics, McGill University, Montr\'eal, Qu\'ebec H3A 2T8, Canada\\
	$^2$ Institut quantique and D\'epartement de physique, Universit\'e de Sherbrooke, 2500 boulevard de l'Universit\'e, Sherbrooke, Qu\'ebec J1K 2R1, Canada\\
	$^3$ Canadian Institute for Advanced Research, Toronto, Ontario M5G 1Z8, Canada}
\date{\today}

\begin{abstract}
 Quantum state transfer into a memory, state shuttling over long distances via a quantum bus, and high-fidelity readout are important tasks for quantum technology. 
 Realizing these tasks is challenging in the presence of realistic couplings to an environment.
 Here, we introduce and assess protocols that can be used in cavity QED to perform high-fidelity quantum state transfer and fast quantum nondemolition qubit readout through Hamiltonian engineering.  
 We show that high-fidelity state transfer between a cavity and a single qubit can be performed, even in the limit of strong dephasing due to inhomogeneous broadening. 
 We generalize this result to state transfer between a cavity and a logical qubit encoded in a collective mode of a large ensemble of $N$ physical qubits. 
 Under a decoupling sequence, we show that inhomogeneity in the ensemble couples two collective bright states to only two other collective modes, leaving the remaining $N-3$ single-excitation states dark.
 Moreover, we show that large signal-to-noise and high single-shot fidelity can be achieved in a cavity-based qubit readout, even in the weak-coupling limit.  These ideas may be important for novel systems coupling single spins to a microwave cavity.  
\end{abstract}


\maketitle

\section{Introduction}
Spin qubits encoded in collective modes of ensembles~\cite{schuster2010high,amsuss2011cavity,kubo2011hybrid} and single spins in quantum dots~\cite{petersson2012circuit,viennot2015coherent,beaudoin2016coupling} can be coupled to microwave cavities for cavity quantum electrodynamics (QED) experiments~\cite{haroche2006exploring}.  Spin qubits show promise for use as long-lived quantum memories, but often suffer from weak qubit-cavity coupling relative to the inhomogeneously broadened linewidth~\cite{putz2017spectral}.
Inhomogeneous broadening typically originates from nuclear-spin or electrical (charge) noise~\cite{balasubramanian2009ultralong,muhonen2014storing,siyushev2014coherent,veldhorst2014an}.
While nuclear-spin noise~\cite{bluhm2011dephasing} can often be controlled through isotopic purification, strong coupling of a single spin to the electric field of a cavity mode typically requires a strong correlation of spin and charge degrees of freedom~\cite{trif2008spin,hu2012strong,jin2012strong,kloeffel2013circuit,tosi2015silicon}. This correlation makes the spin qubit susceptible to low-frequency charge noise~\cite{dial2013charge,laird2013valley}.
An alternative strategy is to enhance the weak magnetic coupling of a spin qubit (which may be otherwise insensitive to charge noise) by coupling to the collective mode of a large spin ensemble~\cite{wesenberg2009quantum,kubo2010strong}. However, spatial inhomogeneities in such ensembles can result in an inhomogeneous linewidth that is comparable to the qubit-cavity coupling~\cite{kubo2011hybrid,kubo2012storage}. 

It is well known that the effects of inhomogeneous broadening can be eliminated through a suitable dynamical decoupling sequence. To determine the quality of a cavity-QED scheme, the coupling is therefore often compared with the inverse qubit coherence time under a train of decoupling $\pi$-pulses~\cite{trif2008spin,hu2012strong,jin2012strong}, rather than the inhomogeneous linewidth. 
However, for a qubit coupled to a cavity, a sequence of $\pi$-pulses typically generates unwanted cavity excitations on the same timescale as coherent qubit-cavity (vacuum Rabi) oscillations, severely reducing the fidelity of, e.g., quantum state transfer between a qubit and a cavity.

In this paper, we show that these limitations can be overcome by engineering appropriate time-averaged Hamiltonians~\cite{viola1999dynamical,khodjasteh2009dynamical,khodjasteh2009dynamically,west2010high} through a combination of qubit dynamical decoupling and control of the qubit-cavity coupling. In particular, we introduce and quantitatively characterize protocols for a high-fidelity quantum state transfer between a qubit and cavity, and for a fast quantum nondemolition qubit readout.
Our readout protocol yields a large signal-to-noise ratio even in the weak-coupling regime, in which the qubit-cavity coupling is small compared to the cavity damping rate. Moreover, we show that control of the qubit-cavity coupling makes high-fidelity quantum state transfer possible even in the strong-dephasing limit, in which the inhomogeneous linewidth dominates the qubit-cavity coupling. This result applies even to logical qubits encoded in the collective mode of an ensemble of physical qubits (relevant to, e.g., spin or atomic ensembles that are routinely used for quantum memories~\cite{kubo2011hybrid,jobez2015coherent,saglamyurek2015quantum}). Inhomogeneous broadening across an uncontrolled ensemble of $N$ physical qubits would typically lead to coupling of the logical qubit to $\sim N$ collective modes~\cite{kubo2012storage,diniz2011strongly,kurucz2011spectroscopic}.  However, remarkably, for our pulse sequence we find that the leading corrections in average Hamiltonian theory couple only four distinct collective modes in the large-$N$ limit. This may allow for very high-fidelity storage-and-retrieval or even coherent manipulation of quantum information in the ensemble through revivals.

This paper is organized as follows. In Section~\ref{secHam}, we introduce the Hamiltonian engineering protocol studied throughout this work. In Section~\ref{secTransfer}, we evaluate the fidelity of a quantum state transfer between a cavity and a single physical qubit under the Hamiltonian-engineering protocol presented here, and show that errors can be strongly suppressed, even in the strong-dephasing limit (in which the inhomogeneous broadening is larger than the qubit-cavity coupling). In Section~\ref{secEnsembles}, we generalize this result to state transfer between a cavity and a collective mode of a large ensemble of physical qubits. 
In Section~\ref{secLimitations}, we analyze realistic control limitations. We focus on finite off/on ratio of the tunable qubit-cavity coupling, deterministic over (under)-rotations of the qubit during imperfect $\pi$-pulses, finite bandwidth of the qubit-cavity coupling pulses, and finite duration of the qubit $\pi$-pulses.
Finally, in Section~\ref{secReadoutMain}, we assess a readout protocol based on the Carr-Purcell sequence that yields high signal-to-noise and single-shot fidelity in the weak-coupling regime.

\section{Hamiltonian engineering \label{secHam}}

We first consider a single qubit coupled to a cavity. With the cavity and the qubit on resonance and working in a rotating frame within the rotating-wave approximation, the system is described by a Jaynes-Cummings Hamiltonian:
\begin{equation}
 H_\mrm{JC}(t)=\xi \sigma_z/2+g(t)(a^\dagger\sigma_- + a\s_+),
\end{equation}
where we have allowed for a tunable qubit-cavity coupling $g(t)$ (setting $\hbar=1$). In addition, the qubit is controlled via $H_\mrm c(t)$, giving the total Hamiltonian
\begin{equation}
 H(t)=H_\mrm{JC}(t)+H_\mrm c(t).
\end{equation}
In $H_\mrm{JC}(t)$, we take $\xi$ to be a Gaussian random variable with zero mean and variance $(\Delta\xi)^2$ that describes inhomogeneous broadening in the qubit-cavity detuning. Most decoupling schemes rely entirely on qubit control. However, electrical control of $g(t)$ is now possible in several architectures~\cite{viennot2015coherent,beaudoin2016coupling,jin2012strong,cottet2010spin,gambetta2011superconducting,srinivasan2011tunable}. By modulating $H_\mrm c(t)$ and $g(t)$ sufficiently quickly, we can eliminate unwanted terms and generate useful time-averaged Hamiltonians. 

\begin{figure}[t]
 \begin{center}
 \includegraphics[width=0.9\textwidth]{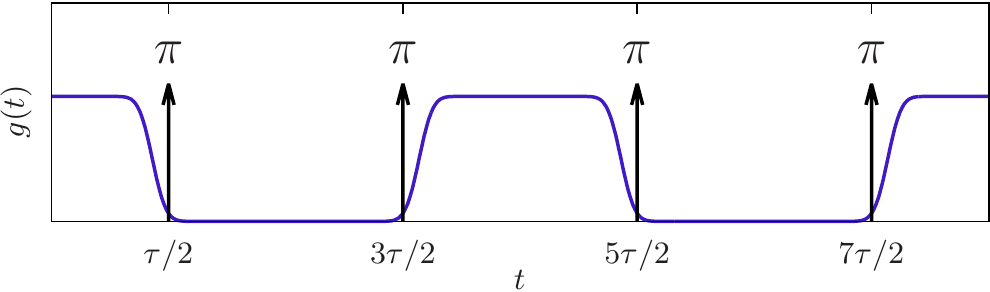}
 \end{center}
 \vspace{-5mm}
 \caption{SQUADD (SQUare wave And Dynamical Decoupling): the Carr-Purcell sequence is applied to a qubit coupled to a cavity while turning off the coupling $g(t)$ [Eq.~\eq{eqnHtoggling}] after each odd-numbered $\pi$-pulse to prevent unwanted cavity excitations.
 \label{figSequence}}
\end{figure}

To average away unwanted terms, we move to the toggling frame~\cite{mehring1976high}, which incorporates $H_\mrm c(t)$ into the transformed system Hamiltonian, 
\begin{align}
 H_\mrm T(t) = U^\dag_\mrm c(t)H_\mrm {JC}(t)U_\mrm c(t),	\label{eqnDefHT}
\end{align}
where $U_\mrm c(t)=\mathcal T \exp[-i\int_0^t dt' H_\mrm c(t')]$.
To reduce dephasing due to the random detuning $\xi$, a natural choice for $U_\mrm c(t)$ is the Carr-Purcell sequence: a train of sharp $\pi$-pulses applied at times $(m+\frac12)\tau$, with $m\,\in\,\mathbb N$ (Fig.~\ref{figSequence}). 
When $U_\mathrm{c}(t)$ generates fast $\pi$-rotations about the $x$-axis,
\begin{align}
 H_\mrm T(t)=\left\{
 \begin{array}{ll}
 \textstyle{\frac12}\xi\s_z+g(t)[a^\dag\s_-+a\s_+],	&n(t)\;\mbox{even},\\
 \textstyle{-\frac12}\xi\s_z+g(t)[a^\dag\s_++a\s_-],	&n(t)\;\mbox{odd},
 \end{array}
 \right.	\label{eqnHtoggling}
\end{align}
with $n(t)$ the number of $\pi$-pulses applied before time $t$. 
For $n(t)$ even, the qubit-cavity interaction is described by a co-rotating term, preserving the total number of excitations, $N_\mrm{ex}\equiv a^\dag a+\s_+\s_-$. However, for $n(t)$ odd, the interaction is rather given by a counter-rotating term, which does not conserve $N_\mrm{ex}$.  For fixed $g(t)=g$, the counter-rotating term leads to simultaneous excitation of the qubit and cavity. This flow of excitations can be blocked simply by taking $g(t)=0$ for $n(t)$ odd. With this choice, $N_\mrm{ex}$ is a constant of motion, allowing for coherent state transfer (vacuum Rabi oscillations) between the qubit and the Hilbert space spanned by the vacuum and first excited state of the cavity. 

In the rest of this paper, we will use the acronym SQUADD (SQUare wave And Dynamical Decoupling) to describe the simultaneous square-wave modulation of $g(t)$ and sequence of $\pi$-pulses shown in Fig.~\ref{figSequence}. In an ideal implementation of SQUADD, qubit $\pi$-pulses are infinitely narrow and coupling modulations perfectly square: $g(t)=g$ for $n(t)$ even and $g(t)=0$ for $n(t)$ odd.  We focus on this idealized limit in Sections~\ref{secTransfer} and \ref{secEnsembles}.  Control imperfections will be considered in detail in Section~\ref{secLimitations}.  In Section~\ref{secReadoutMain}, we will show that the Carr-Purcell sequence with fixed $g(t)=g$ can be used for fast readout of the qubit via the cavity.

\section{Qubit-cavity state transfer \label{secTransfer}}
In this section, we assess the quality of a quantum state transfer realized using SQUADD. In particular, we show that SQUADD allows for a high-fidelity quantum state transfer, even in the limit of strong inhomogeneous broadening, $\D\xi>g$. 

To characterize the performance of SQUADD, we evaluate the average fidelity
\begin{align}
 F = \int d\psi\bra \psi U_0^\dag\mathcal M(\proj\psi)U_0\ket\psi.	\label{eqnDefF}
\end{align}
The integral in Eq.~\eq{eqnDefF} represents an average with respect to the Haar measure $d\y$ (a uniform average over the Bloch sphere) for the ensemble of states of the form $\ket\y\equiv\ket\y_\mrm q\ket 0_\mrm c$, where $\ket \y_\mrm q$ is a pure qubit state and $\ket 0_\mrm c$ is the cavity vacuum. We have also introduced the unitary operator $U_0$ describing an ideal state transfer: $U_0\ket\y_\mrm q\ket 0_\mrm c=\ket g_\mrm q\ket\y_\mrm c$, with $\ket g_\mrm q$ the qubit ground state. In addition, $\mathcal M$ is the completely positive trace-preserving map that describes the actual state transfer, accounting for an average over the random detuning $\xi$ and a finite cavity damping rate $\kappa$. We first consider the case $\kappa=0$, then generalize to finite $\kappa$, below.  

\subsection{Exact solution ($\kappa=0$)}

\begin{figure}
 \begin{center}
 \includegraphics[width=0.9\textwidth]{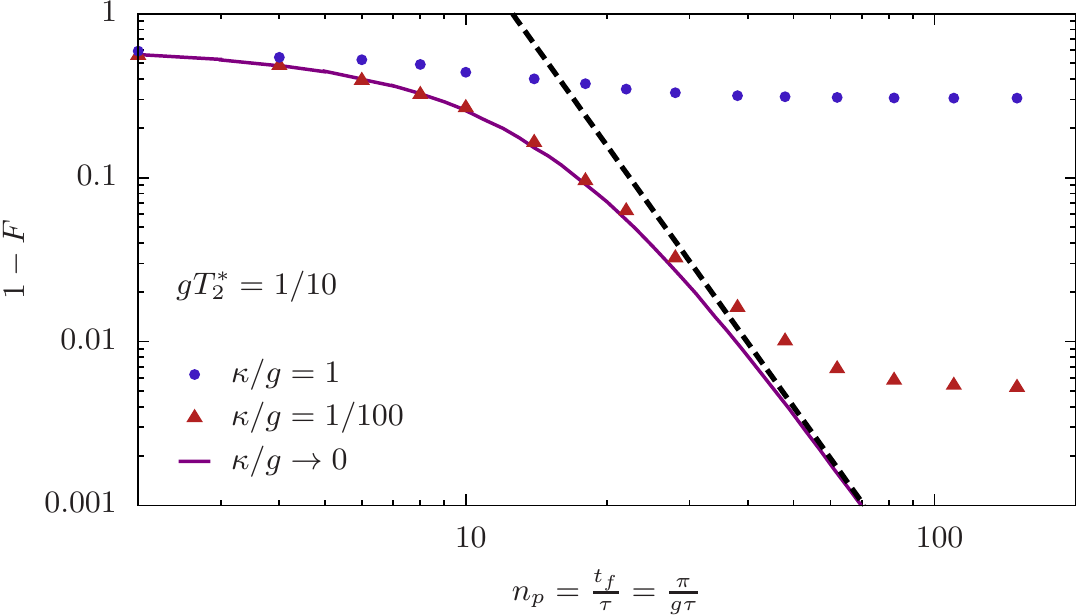}
 \end{center}
 \vspace{-5mm}
 \caption{Suppression of state-transfer error $1-F$ with increasing number of pulses $n_p$ for $gT_2^\ast=\sqrt{2}g/\D\xi=1/10$. Dashed black line: Eq.~\eq{eqnFid}. Solid purple line: exact solution without cavity damping, Eq.~\eq{eqnFexacte}. Blue dots: exact numerical master-equation simulation including cavity damping, with $\kp/g=1$. Red triangles: $\kp/g=1/100$. 
 \label{figQS}}
\end{figure}

An exact solution is possible for SQUADD under the ideal conditions described above: sharp $\pi$-pulses, $g(t)=g$ for $n(t)$ even, and $g(t)=0$ for $n(t)$ odd. The time-evolution operator then breaks into segments associated with the intervals of duration $\tau$ between $\pi$-pulses. In the single-excitation subspace, $N_\mrm{ex}=a^\dag a+\s_+\s_-=1$, the evolution operator for a single period of the decoupling sequence is
\begin{align}
 U_1 = R_{\hat{n}} (\W\tau)R_{\hat{z}}(-\xi\tau)R_{\hat{n}}(\W\tau),\label{eqnU1}
\end{align}
with
\begin{align}
 \W&=\sqrt{g^2+\xi^2/4},\\
 \hat{n} &= \frac g\W\hat{x}+\frac\xi{2\W}\hat{z}.
\end{align}
In Eq.~\eq{eqnU1}, we have introduced the operator $R_{\hat{n}}(\q)\equiv\eul{-i\q \hat{n}\cdot\mvec\tau/2}$, which applies an SU(2) rotation by angle $\q$ around the axis set by the unit vector $\hat{n}$ in the space spanned by the vector of pseudospins $\mvec \tau=(\tau_x,\tau_y,\tau_z)$. These pseudospins are defined by
\begin{align}
 \tau_x&=\ket{g1}\bra{e0}+\ket{e0}\bra{g1},	\label{eqntx}\\
 \tau_y&=i(\ket{g1}\bra{e0}-\ket{e0}\bra{g1}),\\
 \tau_z&=\proj{e0}-\proj{g1}.	\label{eqntz}
\end{align}
In Eqs.~\eq{eqntx} to \eq{eqntz}, $g$ ($e$) labels the ground (excited) state of the qubit, while $0$ or $1$ is the number of photons in the cavity. The product of the three rotation matrices in Eq.~\eq{eqnU1} is itself a rotation matrix $U_1=R_{\hat v}(\vtt)$, where
\begin{align}
 \cos\frac\vtt2 &= \sqrt{1-A^2-B^2},	\label{eqnTheta}\\
 \hat{v} &=\frac{A}{\sqrt{A^2+B^2}}\hat{x}+\frac{B}{\sqrt{A^2+B^2}}\hat{z},
 \label{eqnv}
\end{align}
with
\begin{align}
 A &= \frac{2g}\W\left(\cos\frac{\W\tau}2\cos\frac{\xi\tau}2 + \frac\xi{2\W}\sin\frac{\W\tau}2\sin\frac{\xi\tau}2\right)\sin\frac{\W\tau}2,	\label{eqnA}\\
 B &=\left(\frac\xi\W\sin\frac{\W\tau}2\cos\frac{\xi\tau}2 -  
    \cos\frac{\W\tau}2\sin\frac{\xi\tau}2 \right)\cos\frac{\W\tau}2
    +\frac{\xi^2-4g^2}{4\W^2}\sin^2\frac{\W\tau}2\sin\frac{\xi\tau}2.	\label{eqnB}
\end{align}
The evolution at the end of the full sequence of $n_p$ pulses (and thus $n_p/2$ periods) is given by
\begin{align}
 U(t_f)=U_1^{n_p/2}=R_{\hat{v}}(n_p\vtt/2),	\hspace{2cm} (n_p\;\mrm{even}).	\label{eqnSolExacte}
\end{align}
Equation \eq{eqnSolExacte} gives a closed-form analytical expression for the evolution operator under SQUADD. 
Taking $\mathcal M(\proj\y)=U(t_f)\proj\y U^\dag(t_f)$ in Eq.~\eq{eqnDefF}, with $U(t_f)$ given by Eq.~\eq{eqnSolExacte}, we obtain the average state-transfer fidelity
\begin{align}
 F=\frac13\mrm E\left[1+v_x^2\sin^2\frac{n_p\vtt}{4}+v_x\sin\frac{n_p\vtt}4\right]	\label{eqnFexacte},
\end{align}
where $\mrm E[\cdot]$ is an ensemble average over the detuning $\xi$.  Equation \eqref{eqnFexacte} is plotted as a function of $n_p$ in Fig.~\ref{figQS} (purple solid line).

To clarify the parametric dependences in Eq.~\eqref{eqnFexacte}, we set $\tau=\pi/(g n_p)$ for a complete state transfer [minimizing error to leading order in $\max(g,\xi)\tau$] and expand to leading order in $1/n_p$.  This gives
\begin{align}
 &1-F\simeq\frac16\left[\left(\frac \pi 4\right)^2\left(\frac{\D\xi}g\right)^4+\frac13\left(\frac{\D\xi}g\right)^2\right]\left(\frac\pi{2n_p}\right)^4,	\label{eqnFid}
\end{align}
valid for $\max(g,\D\xi)\tau\ll1$ [equivalently, $n_p\gg\pi\max(1,\D\xi/g)$]. The error ($1-F\propto 1/n_p^4$) is thus strongly suppressed with an increasing number of $\pi$-pulses, as shown in Fig.~\ref{figQS}. 

A small error can be reached when $n_p\gg1$, even for strong dephasing, $gT_2^*=\sqrt{2}g/\D\xi<1$, where $T_2^*$ is the qubit free-induction decay time (dephasing time) due to inhomogeneous broadening $\Delta\xi$. This result is apparent in Fig.~\ref{figQS}, which gives the exact solution described above (solid purple line), along with the large-$n_p$ expansion of Eq.~\eq{eqnFid} (dashed black line). Here, we have chosen $g T_2^\ast=1/{10}$. Even for this choice of parameters, placing the system in the strong-dephasing regime, errors smaller than $1\%$ are reached with $n_p\sim40$ pulses, at the onset of the validity criterion for Eq.~\eq{eqnFid}: $n_p>\pi\D\xi/g\sim40$. Consequently, the usual weak-dephasing criterion ($1/T_2^\ast\ll g$) has been traded for a fast-control requirement ($\tau\ll T_2^\ast$)~\footnote{Here, we consider static noise $\xi$, corresponding to an infinite correlation time, $\tau_c\rightarrow\infty$. For a fluctuating detuning with a finite correlation time, the pulse interval $\tau$ must also satisfy $\tau<\tau_c$ for significant error suppression~\cite{khodjasteh2011limits}.}. Fast $\pi$-pulses in this limit have already been demonstrated with isolated spin qubits (not coupled to cavities)~\cite{bluhm2011dephasing,dial2013charge,yoneda2014fast}, and could in principle be made even faster for single spins by taking advantage of exchange coupling and the magnetic field gradient generated by a micromagnet~\cite{coish2007exchange,chesi2014single}. Since $g(t)$ can be controlled electrically when these systems are coupled to cavities~\cite{viennot2015coherent,beaudoin2016coupling}, fast pulsing of $g(t)$ may be possible in the very near future (we give an analysis of finite-bandwidth control for $g(t)$ and the influence of counter-rotating terms in Section~\ref{secBandePassante}).

\subsection{Finite cavity damping ($\kappa\ne 0$)}

When $n_p\to\infty$, inhomogeneous broadening becomes irrelevant and the fidelity will ultimately be limited by cavity damping at rate $\kp$ (we neglect intrinsic qubit decay due to a homogeneous linewidth when $\kappa T_2>1$). Accounting for finite cavity damping $\kappa\ne 0$ in the map $\mathcal{M}$ in Eq.~\eqref{eqnDefF}, and expanding for $\kappa/g\ll 1$, we find that the error saturates at 
\begin{align}
 1-F = \frac\pi6\,\frac\kp g+\mathcal O\left(\frac{\kp^2}{g^2}\right)	\label{eqnFidKappa}
\end{align}
when $n_p\rightarrow\infty$. To establish the influence of cavity damping more generally as a function of $\kappa/g$ and $n_p$, we numerically solve the Lindblad master equation generated by a Liouvillian superoperator $\liouv$ accounting for both  Hamiltonian evolution under Eq.~\eq{eqnHtoggling} and cavity damping. As shown in Fig.~\ref{figQS}, cavity damping does indeed lead to a saturation of the error as a function of $n_p$ at $1-F\sim\kp/g$ (blue dots: $\kp/g=1$, red triangles: $\kp/g=1/100$). 

As a concrete example, a coupling $g/2\pi\simeq1$~MHz has been predicted for spin qubits in GaAs double quantum dots~\cite{hu2012strong}, leading to $gT_2^\ast\simeq0.05$ due to hyperfine coupling to nuclear spins~\cite{bluhm2011dephasing}. Even in this case, SQUADD could enable coherent coupling between a single spin and a cavity. In addition, SQUADD could improve state transfer between a single spin confined in a carbon nanotube and a coplanar-waveguide resonator. In a recent experiment on this system, $g/2\pi=1.3$~MHz, $\kp/2\pi=0.6$~MHz, and $T_2^\ast\simeq 60$~ns have been reported~\cite{viennot2015coherent}. With these parameters, a large state-transfer error $1-F\simeq0.42$ results from Eq.~\eqref{eqnDefF} without $\pi$-pulses. Using SQUADD, $n_p=10$ ($\tau=\frac\pi{gn_p}\simeq40$~ns) suffices to reduce the error from pure dephasing to $0.004$.
The total error is then $1-F\simeq0.18$, limited by the large $\kp/g$ ratio in this experiment.

\subsection{Average Hamiltonian theory\label{secAHT}}

For $\kp=0$, some insight into the dependence of the state-transfer error on $n_p$ can be gained using average Hamiltonian theory, a standard tool in the analysis of open-loop Hamiltonian engineering protocols~\cite{viola1999dynamical}. In average Hamiltonian theory, the evolution operator $U(t)=\mathcal T\exp[-i\int_0^tdt'H_\mrm T(t')]$ is recast in terms of a Magnus expansion \cite{haeberlen1968coherent,blanes2009magnus}:
\begin{equation}
 U(t)=\exp\left[-it\sum_{k=0}^\infty \Hb{k}\right].	\label{eqnHavg}
\end{equation}
Substituting $H_\mrm T(t)$ [Eq.~\eq{eqnHtoggling}] into the expressions for $\Hb{k}$~\cite{haeberlen1968coherent,blanes2009magnus} gives the first few terms in the Magnus expansion for SQUADD
\begin{align}
 \Hb{0}&=\frac1{2\tau}\int_0^{2\tau}dt_1H_\mrm T(t_1)=\frac g2(a^\dag\s_-+a\s_+), \label{eqnHmoyen0}\\
 \Hb{1}&=-\frac i{4\tau}\int_0^{2\tau}dt_1\int_0^{t_1}dt_2\left[H_\mrm T(t_1),H_\mrm T(t_2)\right]=0,	\label{eqnHmoyen1}\\
 \Hb{2}&=-\frac 1{12\tau}\int_0^{2\tau}dt_1\int_0^{t_1}dt_2\int_0^{t_2}dt_3
	    \Big\{\left[H_\mrm T(t_1),\left[H_\mrm T(t_2),H_\mrm T(t_3)\right]\right]\notag\\
		&\hspace{6cm}+\left[\left[H_\mrm T(t_1),H_\mrm T(t_2)\right],H_\mrm T(t_3)\right]
	    \Big\}\notag\\
  &=-\frac{g\xi^2\tau^2}{48}(a\s_++a^\dag\s_-) -\frac{g^2\xi\tau^2}{24}\left(a^\dag a+\frac12\right)\s_z.	\label{eqnHmoyen2}
\end{align}
The leading term, $\Hb{0}$, generates coherent vacuum Rabi oscillations between the qubit and the cavity. In Eq.~\eq{eqnHmoyen1}, $\Hb{1}$ vanishes because the toggling-frame Hamiltonian given by Eq.~\eq{eqnHtoggling} has the following property: $H_\mrm T (t) = H_\mrm T (2\tau - t)$, corresponding to a symmetric cycle with period $T=2\tau$~\cite{mehring1976high}. For such symmetric cycles, all odd orders vanish in the Magnus expansion, leading to $\Hb{1} = 0$~\cite{wang1972decay}. The leading source of error is then $\Hb{2}$. In the subspace containing a single qubit or cavity excitation ($N_\mathrm{ex}=1$), the first and second terms in $\Hb{2}$ [Eq.~\eq{eqnHmoyen2}] lead to unwanted rotations by an angle $\propto\tau^2$ around the axes defined by $\tau_x$ [Eq.~\eq{eqntx}] and $\tau_z$ [Eq.~\eq{eqntz}], respectively. For small rotation angles, this gives a correction $\propto (\tau^2)^2$ to the fidelity (which involves an overlap between two state vectors at an angle $\propto \tau^2$), thus explaining the error $\propto \tau^4\propto 1/n_p^4$ obtained in Eq.~\eq{eqnFid}. 

A sufficient criterion for convergence of the Magnus expansion [Eq.~\eq{eqnHavg}] for a periodic Hamiltonian $H_\mathrm{T}(t)$ with period $2\tau$ is~\cite{blanes2009magnus}
\begin{align}
 \int_0^{2\tau}\!dt\|H_\mrm T(t)\|<\pi,	\label{eqnConvergenceMagnus}
\end{align}
where $\|O\|$ is the 2-norm of $O$, with $O$ an arbitrary operator.  Here, we have taken the detuning $\xi$ to be Gaussian distributed over an infinite interval and the cavity mode is described by unbounded operators $a,a^\dagger$.  Thus, $\|H_\mrm T(t)\|$ is formally unbounded.  However, realistically, we expect that the Gaussian distribution for $\xi$ can be truncated at a few $\Delta\xi$, and the cavity occupation can be truncated to include only a few quanta in Fock space, resulting in $||a||\sim||a^\dagger||\sim 1$, which gives $\int_0^{2\tau}\!dt\|H_\mrm T(t)\|\lesssim\mathrm{max}(g,\Delta\xi)2\tau$ up to factors of order unity.  Thus, for a fixed transfer time $t_f=n_p\tau=\pi/g$, we expect convergence of the Magnus expansion for $n_p\gtrsim2\max(1,\D\xi/g)$. 

Absolute convergence of the Magnus expansion does not guarantee that the first few terms of average Hamiltonian theory lead to an accurate description of the evolution operator. Since the terms of average Hamiltonian theory appear within the argument of an exponential in Eq.~\eq{eqnHavg}, correction terms that are small over a single period $2\tau$ may add up to produce large deviations over the entire evolution time $t_f = n_\mathrm{p}\tau\gg \tau$. However, for SQUADD, the duration of the sequence is set to $t_f=\pi/g$ through the condition for complete state transfer. This sets a bound on the magnitude of deviations of a truncated Magnus expansion from the exact solution. Indeed, for $k\geq2$, we have $\|\Hb{k}\| t_f\lesssim\max(g\D\xi^k,g^k\D\xi)\tau^k (\pi/g)$ $\sim\max[(\D\xi/g)^k,\D\xi/g]/n_p^k$, where we have used $\tau=\pi/g n_p$. Keeping only the first correction term in the Magnus expansion will then lead to an accurate description of the evolution operator (with $\|\Hb k\| t_f\ll1\;\forall\; k\geq2$) in the limit
\begin{align}
 n_p\gg\max\left(1,\frac{\D\xi}g\right).	\label{eqnConditionExacte}
\end{align}
In this limit, Eq.~\eq{eqnConvergenceMagnus} already ensures that the expansion of average Hamiltonian theory is convergent. In this same limit, we find that the expansion given in Eqs.~\eq{eqnHmoyen0} to \eq{eqnHmoyen2} leads to the error given in Eq.~\eq{eqnFid}, thus coinciding with the exact solution.

\section{Collective modes in qubit ensembles	\label{secEnsembles}}

In this section, we consider the application of SQUADD to quantum state transfer between a cavity and a collective mode of an ensemble of $N$ physical qubits. We account for leading corrections in the Magnus expansion and show that, up to corrections $\sim\mathcal O(1/\sqrt N)$, this system evolves in a closed 4-dimensional subspace. Using this approach, we find an expression similar to Eq.~\eq{eqnFid} for the state-transfer fidelity, but applicable to a collective mode.

For single spins coupled to microwave cavities, the $\kappa/g$ ratio can be large~\cite{viennot2015coherent,tosi2014circuit}, limiting the fidelity achievable through SQUADD. However, the effective coupling can be significantly enhanced by encoding a logical qubit into a large number of physical qubits. Indeed, an ensemble of $N$ qubits coupled to a common cavity mode hosts an excitation out of the ground state $\left|g\right>_q=\left|g\right>_1\otimes\left| g \right>_2\cdots \left|g\right>_N$ that is annihilated by the collective lowering operator 
\begin{align}
 b=\sum_{i=1}^N \frac{g_i}{\sqrt{N}g_\mrm{av}}\s_i^-,
 \hspace{13mm}
 g_\mrm{av}\equiv\sqrt{\sum_{i=1}^N\frac{g_i^2}{N}},	\label{eqnb}
\end{align}
where $g_i$ is the coupling for qubit $i$. For $N\gg1$, the logical qubit encoded in the subspace of $\left|g\right>_q$ and $\left|e\right>_q=b^\dagger\left|g\right>_q$ couples to the resonator with an ensemble coupling $g_\mrm{ens}\equiv\sqrt N g_\mrm{av}$~\cite{wesenberg2009quantum}.
However, an inhomogeneity in qubit-cavity detunings across the ensemble may lead to leakage from the collective mode $b$ to many dark states~\cite{kubo2012storage,diniz2011strongly,kurucz2011spectroscopic}.
When $\D\xi\gtrsim g_\mrm{ens}$, leakage due to dephasing will typically result in an error of order one.

Errors due to inhomogeneous broadening in an ensemble can be suppressed through SQUADD. The toggling-frame Hamiltonian for a qubit ensemble is 
\begin{align}
 H_\mrm T(t)=\left\{
 \begin{array}{ll}
  \frac12\sum_i\xi_i\s^z_i + \sum_i g_i(a^\dag \s^-_i+a \s^+_i),	& n(t)\;\mrm{even},	\\
  -\frac12\sum_i\xi_i\s^z_i,	& n(t)\;\mrm{odd}.	\label{eqnHTmain}
 \end{array}
 \right.
\end{align}
We thus consider an ensemble of qubits with couplings $g_i(t)$ and detunings $\xi_i$ from the cavity. As in the single-qubit case, we assume that $g_i(t)=g_i\,\forall\,i$ for $n(t)$ even and $g_i(t)=0\,\forall\,i$ for $n(t)$ odd. The time-dependent Hamiltonian in Eq.~\eq{eqnHTmain} describes rapid periodic modulation for small pulse interval $\tau$. 
Under the conditions described in Section \ref{secAHT}, above, we then expect convergence of the Magnus expansion for $\mathrm{max}(g_\mrm{ens},\D\xi)2\tau\lesssim\pi$.
Because we have assumed $g(t)=0$ for $n(t)$ odd, the total number of excitations $N_\mrm{ex}=a^\dag a + \sum_i\s^+_i\s^-_i$ is a constant of motion.
We thus project each $\Hb{k}$ into the subspace $\mathcal H_{01}$ associated with $N_\mrm{ex}=0$ or 1. Explicitly, $\mathcal H_{01}$ is spanned by the states $\ket{g}_\mrm q\otimes\ket0_\mrm c$, $\ket g_\mrm q\otimes\ket 1_\mrm c$, and $\ket g_1\otimes\ket g_2\cdots\ket e_j\cdots\ket g_N\otimes\ket 0_\mrm c$, where $\ket{0}_\mathrm{c}$ and $\ket{1}_\mathrm{c}$ label cavity Fock states, and where $\ket g_j$ and $\ket e_j$ label the ground state and excited state of qubit $j$, respectively. We then have
\begin{align}
 \Hb{0}&=\frac\gens2(a^\dag b+ab^\dag),	\label{eqnH0}
 \hspace{1cm}
 \Hb{1}=0,\\
 \Hb{2}&=\W_1(b^\dag c+c^\dag b)+\W_2(a^\dag d+d^\dag a)+\ch a^\dag a.
 \label{eqnH2}
\end{align}
In Eqs.~\eq{eqnH0} and \eq{eqnH2}, we have introduced two new collective qubit lowering operators
\begin{align}
 c&=\frac1{\sqrt N}\sum_i\frac{g_i\xi_i}{(g\xi)_\mrm{av}}\s^-_i,
 \hspace{13mm}
 (g\xi)_\mrm{av} = \sqrt{\sum_i \frac{g_i^2\xi_i^2}N},	\label{eqnc}\\
 d&=\frac1{\sqrt N}\sum_i\frac{g_i\xi_i^2}{(g\xi^2)_\mrm{av}}\s^-_i,
 \hspace{10.25mm}
 (g\xi^2)_\mrm{av} = \sqrt{\sum_i \frac{g_i^2\xi_i^4}N}.	\label{eqnd}
\end{align}
Equation \eq{eqnH2} describes the coupling of modes $a$ and $b$ with the new modes $c$ and $d$ with strengths
\begin{align}
 \W_1= -\frac{N\tau^2}{48}g_\mrm{av}(g\xi)_\mrm{av},\hspace{10mm}
 \W_2= -\frac{\sqrt N\tau^2}{48}(g\xi^2)_\mrm{av}.
\end{align}
In addition, Eq.~\eq{eqnH2} contains a resonator frequency shift 
\begin{align}
 \ch=\frac{\tau^2}{24}\sum_i g_i^2\xi_i.
\end{align}
By construction, after projecting into $\mathcal H_{01}$, the collective qubit operators obey the commutation relations
\begin{align}
 [b,b^\dag]=[c,c^\dag]=[d,d^\dag]=1+\ordre(1/N).
\end{align}
Therefore, in the large-$N$ limit, the Hamiltonian $\Hb{0}+\Hb{2}$ can be expanded in the basis of single-excitation states $\ket{m}\equiv m^\dag(\ket{g}_\mrm q\otimes\ket0_\mrm c)$, where $m\;\in\;\{a,b,c,d\}$. However, this basis is typically non-orthogonal. To see this, we consider the case where the coupling strengths $g_i$ are uncorrelated with the detunings $\xi_i$, implying that, e.g., $(g\xi)_\mrm{av}\rightarrow g_\mrm{av}\xi_\mrm{av}$ for $N\rightarrow\infty$. We also take the distribution of qubit-resonator detunings to be Gaussian with mean $\mrm E[\xi_i]=0$. Projecting into $\mathcal H_{01}$, this gives
\begin{align}
 [b,c^\dag]&=\ordre(1/\sqrt N),
 \hspace{15mm}[c,d^\dag]=\ordre(1/\sqrt N),	\label{eqncommbc}\\
 [b,d^\dag]&=1/\sqrt3+\ordre(1/\sqrt N),
\end{align}
all other relevant commutators between different modes being 0. Though $[b,c^\dag]$ and $[c,d^\dag]$ are suppressed in the large-$N$ limit, $[b,d^\dag]$ always remains of order 1. We thus introduce the overlap
\begin{align}
 s\equiv\braket bd=\bra 0[b,d^\dag]\ket 0=1/\sqrt 3+\ordre(1/\sqrt N). \label{eqns}
\end{align}
To avoid the complications associated with the non-orthogonal basis $\{\ket a,\ket b, \ket c,\ket d\}$, we introduce a new set of single-excitation states $\{\ket{\tilde a}, \ket{\tilde b},\ket{\tilde c}, \ket{\tilde d}\}$, where
\begin{align}
 \ket{\tilde a}&=\ket a,\hspace{51mm}\ket{\tilde c}=\ket c,	\label{eqnHatilde}\\
 \ket{\tilde b}&=-\frac1{\sqrt{2(1-s)}}\ket b+\frac1{\sqrt{2(1-s)}}\ket d,
 \hspace{7mm}
 \ket{\tilde d}=\frac1{\sqrt{2(1+s)}}\ket b+\frac1{\sqrt{2(1+s)}}\ket d.	\label{eqnHdtilde}
\end{align}
The states given in Eq.~\eq{eqnHatilde} and \eq{eqnHdtilde} form an orthonormal basis if we neglect overlaps $\sim\ordre(1/\sqrt N)$. 
Writing a matrix representation of $\Hb{0}+\Hb{2}$ in this basis, we find
\begin{align}
 \Big[\Hb{0}+\Hb{2}\Big]=
 \matquatre{
 0	& \w_-	& 0	& \w_+\\
 \w_-	& 0	& \w'_+	& 0\\
 0	& \w'_+	& 0 	& \w'_-\\
 \w_+	& 0	& \w'_-	& 0
 }[1+\ordre(1/\sqrt N)].	\label{eqnTight}
\end{align}
In Eq.~\eq{eqnTight}, we have introduced couplings between the orthonormal modes $\tilde a$, $\tilde b$, $\tilde c$, and $\tilde d$, given by
\begin{align}
 \w_\pm &= \pm\sqrt{\frac{1\pm s}2}\left[\frac{\gens}2\mp\frac{1}{48}g_\mrm{ens}\,(\xi^2)_\mrm{av}\tau^2\right],
 \hspace{5mm}
 \w'_\pm =\pm\frac{1}{48}\sqrt{\frac{1\mp s}2}g_\mrm{ens}^2\,\xi_\mrm{av}\tau^2.	\label{eqnCouplagesEns}
\end{align}

\begin{figure}
 \begin{center}
 \includegraphics[width=0.9\textwidth]{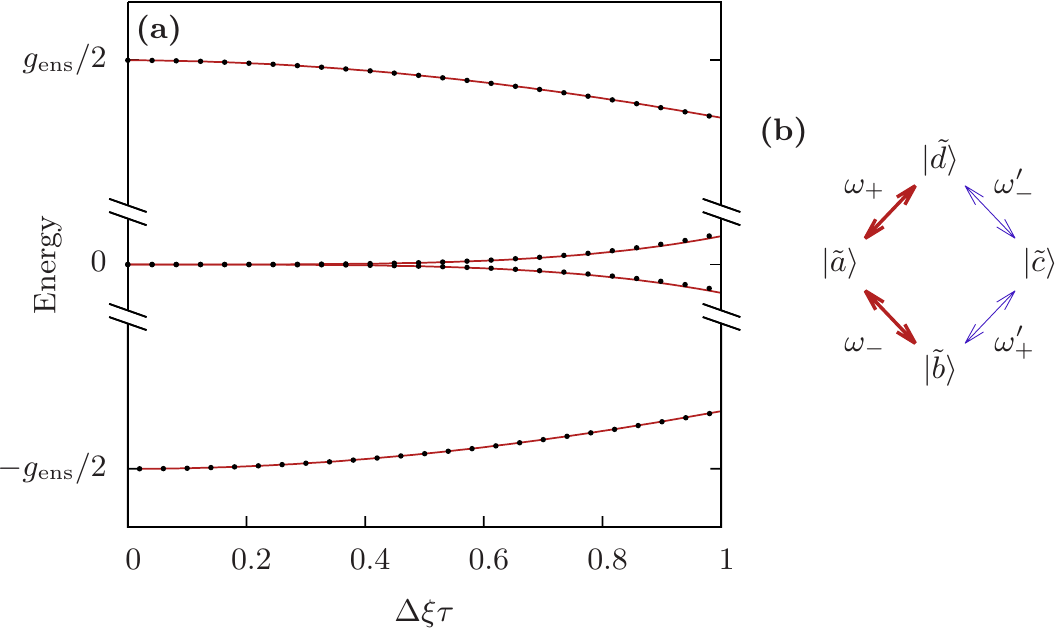}
 \end{center}
 \caption{Spectrum of the time-independent effective Hamiltonian $\Hbc{0}+\Hbc{2}$ for a qubit ensemble coupled to a resonator under SQUADD. (a) Eigenenergies as a function of $\D\xi\tau$, where $\D\xi$ is the standard deviation of the Gaussian distribution of qubit-resonator detunings $\xi_i$, and $\tau$ is the dynamical-decoupling pulse interval. Solid red line: eigenenergies obtained by analytically diagonalizing the effective $4\times4$ Hamiltonian in Eq.~\eq{eqnTight}, dropping corrections $\sim\ordre(1/\sqrt N)$. Black dots: exact numerical diagonalization of $\Hbc{0}+\Hbc{2}$ given by Eqs.~\eq{eqnH0} and \eq{eqnH2}, which include corrections $\sim\ordre(1/\sqrt N)$. We take $\gens=\D\xi$, and $N=1000$. (b) Couplings between the basis states $\{\ket{\tilde a},\ket{\tilde b},\ket{\tilde c},\ket{\tilde d}\}$, as given by Eq.~\eq{eqnTight}.	\label{figdiag}}
\end{figure}

For $N\gg1$, we neglect corrections $\sim \ordre(1/\sqrt N)$. The Hamiltonian in Eq.~\eq{eqnTight} can then be represented graphically, as shown in Fig.~\ref{figdiag}(b). For $\D\xi\tau=0$, Eq.~\eq{eqnCouplagesEns} yields $\w'_\pm=0$. The Hamiltonian then has the structure of a $\Ld$ system, with a basis state $\ket{\tilde a}$ coupled to the two basis states $\ket{\tilde b}$ and $\ket{\tilde d}$. These couplings are represented by the thick red arrows in Fig.~\ref{figdiag}(b). In this limit, the Hamiltonian in Eq.~\eq{eqnTight} has two bright eigenstates with energies $\pm\gens/2$. This is clearly seen in Fig.~\ref{figdiag}(a), which shows the eigenenergies of the Hamiltonian in Eq.~\eq{eqnTight} as a function of $\D\xi\tau$, where $\D\xi$ is the standard deviation of the Gaussian distribution of qubit-resonator detunings. For $\D\xi\tau=0$, all other $N-1$ eigenstates are zero-energy dark states which do not couple to the resonator mode. In contrast, when $\D\xi\tau>0$, the Hamiltonian has the structure of a tight-binding problem on a ring; the basis states $\ket{\tilde b}$ and $\ket{\tilde{d}}$ become weakly coupled through an additional mode $\ket{\tilde c}$. These additional hopping terms are represented by the thin blue arrows in Fig.~\ref{figdiag}(b). 

Because of the simple tridiagonal form of the Hamiltonian in Eq.~\eq{eqnTight}, we are able to find analytic expressions for the eigenenergies of $\Hb{0}+\Hb{2}$ for $\D\xi\tau>0$. Neglecting corrections $\sim\mathcal O(1/\sqrt N)$, diagonalization of the Hamiltonian in Eq.~\eq{eqnTight} reveals two energy doublets: (i) a doublet of bright states (which have finite overlap with the cavity state $\ket{\tilde a}=\ket a=a^\dag\ket0$), and (ii) a doublet of states which are dark (no overlap with $\ket{\tilde a}$) for $\D\xi\tau=0$, but which become bright for $\D\xi\tau>0$. The eigenenergies of these two doublets are
\begin{align}
 E_\pm^\mrm{(i)}=\pm\sqrt{\frac{\w_\mrm{tot}^2+\Sigma^2}2},
 \hspace{1cm}
 E_\pm^\mrm{(ii)}=\pm\sqrt{\frac{\w_\mrm{tot}^2-\Sigma^2}2},	\label{eqnEnergies}
\end{align}
respectively, where
\begin{align}
 \w_\mrm{tot}^2&=\w_+^2+\w_-^2+(\w'_+)^2+(\w'_-)^2,\\
 \Sigma^2&=\sqrt{[(\w_++\w'_+)^2+(\w_--\w'_-)^2][(\w_+-\w'_+)^2+(\w_-+\w'_-)^2]}.
\end{align}
The eigenenergies given in Eq.~\eq{eqnEnergies} are shown in Fig.~\ref{figdiag}(a) (solid red lines). Introducing couplings to the mode $\ket{\tilde c}$ by turning on $\D\xi\tau$ shifts the energies of the two initial bright states, and lifts the degeneracy between two of the initial dark states by coupling them to the resonator mode. These two effects will generate errors in the state transfer of the resonator quantum state into the ensemble of qubits.

We characterize errors in SQUADD due to inhomogeneous broadening with the average fidelity $F$, as defined in Eq.~\eq{eqnDefF}. We consider the initial state $\ket{\psi}\equiv\ket{g}_\mrm q\otimes\ket{\psi}_\mrm c$, where $\ket\psi_\mrm c$ is an arbitrary superposition of the cavity states $\ket{0}_\mrm c$ and $\ket 1_\mrm c$. We choose the evolution operator for an ideal state transfer to be $U_0=-ib^\dag a$, where the $-i$ phase factor appears because the state transfer described here is equivalent to an SU(2) rotation. We take the linear map $\mathcal M$ representing imperfect state transfer to correspond to the evolution operator under the effective time-independent Hamiltonian in Eq.~\eq{eqnTight}. We perform a Taylor expansion of the resulting fidelity to leading (fourth) order in $\tau$, assuming $\max(\gens,\D\xi)\tau\ll1$. Using the condition for complete state transfer, $\tau=\pi/g_\mrm{ens} n_p$, this assumption becomes $n_p\gg\max(\pi,\D\xi/\gens)$, resulting in
\begin{align}
 1-F\simeq\left[\frac{8+\pi^2}{18}\left(\frac{\D\xi}{2\gens}\right)^4+\frac1{18}\left(\frac{\D\xi}{2\gens}\right)^2\right]\left(\frac\pi{2n_p}\right)^4\!\!.	\label{eqnFidEnsemble}
\end{align}
We recall that we have dropped corrections of $\ordre(1/\sqrt N)$ arising from overlaps between basis states.
Ignoring numerical prefactors of order 1, Eq.~\eq{eqnFidEnsemble} exactly corresponds to Eq.~\eq{eqnFid} for a single qubit, after the replacement $g\rightarrow\gens$.
In Eq.~\eq{eqnFidEnsemble}, the numerical prefactors (obtained for $N\gg1$) differ from those obtained in Eq.~\eq{eqnFid} for $N=1$ because the mode structure is not the same. Indeed, taking $N=1$ in Eqs.~\eq{eqnb}, \eq{eqnc}, and \eq{eqnd} leads to $b=c=d=\s_-$. 
The overlap between excitations of any pair of modes $\in\{b,c,d\}$ is then $\bra{0}\s_-\s_+\ket0=1$, in contrast with the overlaps obtained from Eqs.~\eq{eqncommbc} to \eq{eqns} when neglecting terms $\sim\ordre(1/\sqrt N)$ for $N\gg1$.

The above discussion shows that SQUADD is robust to inhomogeneous broadening, even when coupling a cavity to a collective mode. By modulating the detuning rather than the coupling, it may be possible to use a variation of SQUADD on ensembles of nitrogen vacancy (NV)-center spin qubits in diamond coupled to superconducting coplanar waveguides, for which $\D\xi\sim g_\mrm{ens}$ has been reported~\cite{kubo2012storage}. 

This treatment of collective modes in qubit ensembles also demonstrates a clear advantage of our analytical approach over brute-force numerical methods for optimal control. Indeed, the time required for numerical exponentiation of the full system Hamiltonian grows exponentially with ensemble size, making the problem numerically challenging for $N\gg1$. In contrast, the analytical approach reveals a closed 4-dimensional subspace in the same large-$N$ limit.

\section{Control limitations	\label{secLimitations}}

In this section we evaluate the robustness of SQUADD to control imperfections. We first discuss errors due to a finite off/on ratio in the qubit-cavity coupling $g(t)$ and pulse errors due to deterministic over-rotation or under-rotation of the qubit. We then describe the effects of a finite bandwidth and of counter-rotating terms in the qubit-cavity coupling Hamiltonian. We finally discuss error due to finite qubit $\pi$-pulse duration, and show that this source of error can be efficiently mitigated by properly alternating the qubit $\pi$-pulse rotation direction.

\subsection{Finite off/on ratio\label{secOffOn}}

The ideal SQUADD sequence analyzed in Section \ref{secTransfer} assumes that the coupling can be tuned to vanish identically in the ``off'' state. As a consequence, all terms in average Hamiltonian theory [Eqs.~\eq{eqnHmoyen0} to \eq{eqnHmoyen2}] commute with the total number of excitations $N_\mrm{ex}$. If there were some residual coupling $g(t)\neq0$ for $n(t)$ odd, terms that do not commute with $N_\mrm{ex}$ would appear in average Hamiltonian theory. Indeed, $\Hb{0}$ would then contain a term $\propto g(a\s_-+a^\dag\s_+)$. In addition, for a sequence that is not a symmetric cycle [$H_\mrm T (t) \neq H_\mrm T (2\tau - t)$], $\Hb1$ would contain, e.g., a cavity-squeezing term $\propto i g^2\tau(a^2-a^\dag\,\!^2)\s_z$. This squeezing term may be useful, e.g., to enhance the fidelity of a qubit readout~\cite{didier2015fast}.

Given a finite off/on ratio $g_\mrm{off}/g$ [where $g(t)=g_\mrm{off}$ for $n(t)$ odd], the term $\propto g(a\s_-+a^\dag\s_+)$ in $\Hb{0}$ generates a correction to the error given in Eq.~\eqref{eqnFid} of order $\sim (g_\mrm{off}/g)^2$. This correction would ultimately limit the saturation fidelity at large $n_p$ whenever $g_\mrm{off}\ne 0$.

\subsection{Pulse errors\label{secPulses}}

In general, over-rotation or under-rotation of the qubit due to imperfect control can lead to an accumulation of errors as the number of pulses $n_p$ is increased.
A simple way to avoid accumulation of these pulse errors is to use a phase-alternated sequence~\cite{mehring1976high}, in which the qubit rotation direction alternates from one $\pi$-pulse to the next.
Consequently, the (fixed, deterministic) error $\ve$ on the rotation angle of successive pulses cancels for $n(t)$ even, but introduces a small over-rotation for $n(t)$ odd. We evaluate the resulting correction $\dt F$ to the state-transfer fidelity of Eq.~\eq{eqnFid} by expanding $U(t)=\mathcal T\exp[-i\int_0^tdt'H_\mrm T(t')]$ to leading order in $\ve$ and in $\tau$.  When $\mathrm{max}(g,\D\xi)\tau\ll1$, $\dt F$ is then well-approximated by this leading correction: 
\begin{align}
 \dt F\simeq -\frac12\left(\frac43-\frac1{\sqrt 2}\sin\frac\pi{\sqrt2}\right)\left(\ve\frac{\D\xi}{g}\right)^2.
 \label{eqnPulse}
\end{align}
Thus, neglecting order-unity prefactors, pulse errors can be made negligible compared to the error given in Eq.~\eq{eqnFid} when $\ve \ll \mathrm{max}\left[\left(\Delta\xi/g\right)^2,1\right]/n_p^2$.

\subsection{Finite bandwidth and counter-rotating terms	\label{secBandePassante}}

In this section, we evaluate the state-transfer fidelity using numerical simulations that take into account both the finite bandwidth of the coupling modulation $g(t)$ and the counter-rotating terms in the Rabi Hamiltonian. In these simulations, we find the evolution of the system under the toggling-frame Hamiltonian
\begin{align}
 H_\mrm T(t)=\left\{\barr{ll}
    H_\mrm T^\mrm{even}(t),	& n(t)\;\mrm{even},\\
    H_\mrm T^\mrm{odd}(t),	& n(t)\;\mrm{odd}.
 \earr
 \right.	\label{eqnHtoggCR}
\end{align}
where
\begin{align}
 H_\mrm T^\mrm{even}(t)&={\textstyle\frac12}\xi\s_z + g(t)\left[a^\dag\s_-+a^\dag\s_+\eul{2i\w_\mrm q t}+\hc\right],\notag\\
 H_\mrm T^\mrm{odd}(t)&={-\textstyle\frac12}\xi\s_z + g(t)\left[a^\dag\s_++a^\dag\s_-\eul{2i\w_\mrm q t}+\hc\right],
      \label{eqnHTCR}
\end{align}
with $\w_\mrm q$ the qubit frequency (assumed to be tuned to equal the resonator frequency). In contrast with Eq.~\eq{eqnHtoggling}, Eqs.~\eq{eqnHTCR} take into account the counter-rotating terms appearing in the Rabi Hamiltonian. These terms give rise to leakage outside the subspace of zero or one excitation when $g(\w)\equiv\int_{-\infty}^\infty dt\exp[i\w t] g(t)$ has significant weight at $\w=2\w_\mrm q$.

\begin{figure}
 \begin{center}
  \includegraphics[width=0.9\textwidth]{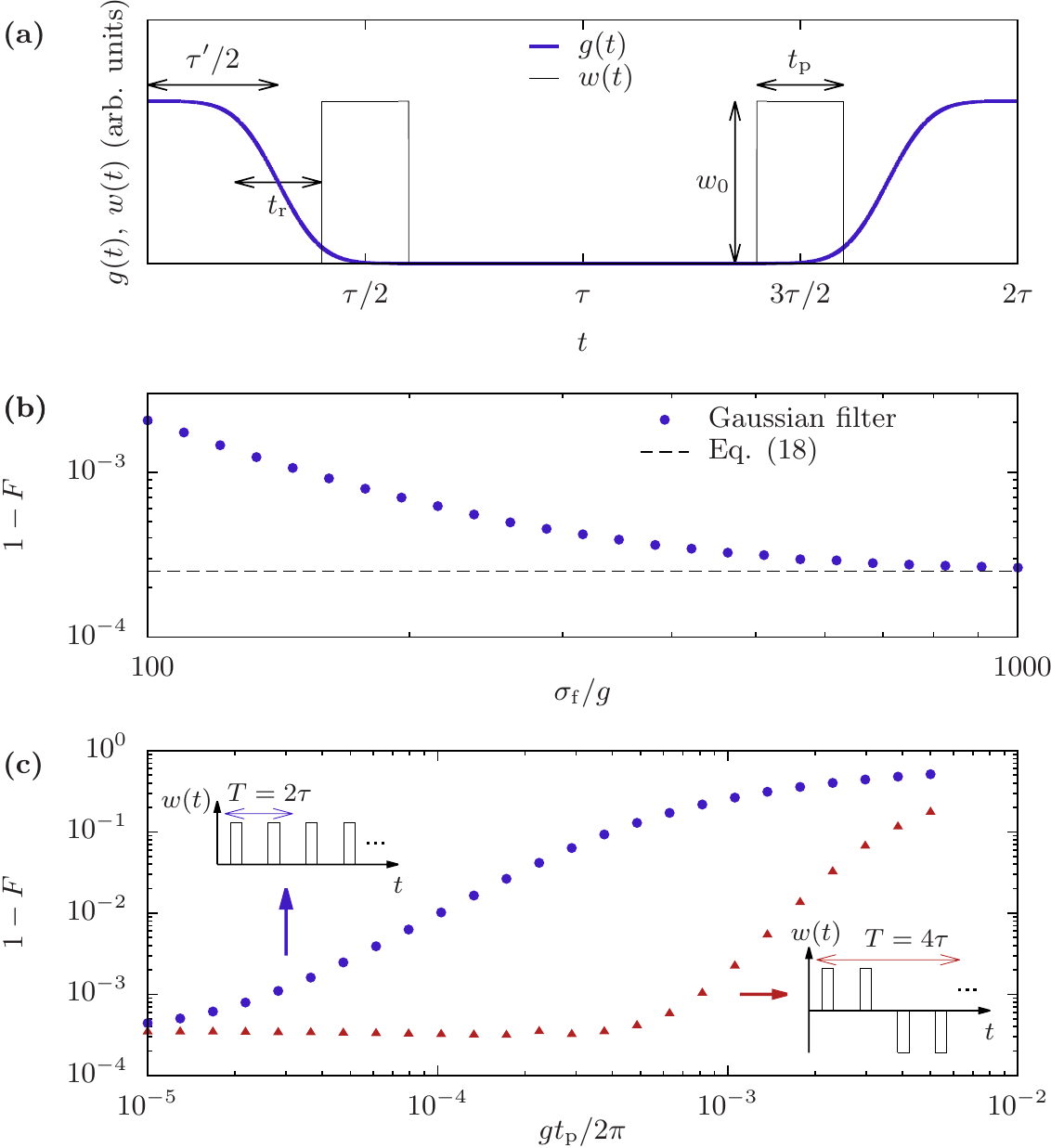}
 \end{center}
 \caption{Error due to control limitations. (a) Coupling $g(t)$ (thick blue line) and amplitude $w(t)$ of the qubit drive (thin black line) for a single period of SQUADD. The coupling is given by Eq.~\eq{eqngdet}, which takes into account a finite bandwidth $\sigma_\mrm f$ leading to a finite rise time $t_\mrm r$, Eq.~\eq{eqntr}. Coupling pulses have a width $\tau'$. Qubit $\pi$-pulses are rectangular with amplitude $w_0$ and duration $t_\mrm p$.
 (b) Error $1-F$ due to finite $t_\mathrm{r}$. Blue dots: error from a finite-bandwidth modulation [resulting from a Gaussian filter with standard deviation $\s_\mrm f$]. Dashed black line: error for a perfectly square modulation of $g(t)$, given by Eq.~\eq{eqnFid}. The parameters used for this plot are: $n_p=100$, $gT_2^\ast=1/10$, $\w_\mrm q=2000g$, $\tau'=\tau-t_\mrm r$, and $\kp=0$. 
 (c) Error under SQUADD using a train of qubit $\pi$-pulses with duration $t_\mrm p$. Blue dots: error from $\pi$-pulses with identical phase (rotations about $+\hat{x}$; top-left inset, period $T=2\tau$). Red triangles: error from $\pi$-pulses that alternate rotations about $\hat{x}$ and $-\hat{x}$ after every pair of pulses (bottom-right inset, period $T=4\tau$). Parameters are $n_p=100$, $gT_2^\ast=1/10$, $\s_\mrm f/g=1000$, $\tau'=\tau-t_\mrm r-t_\mrm p$, and $\kappa=0$.
 \label{figBandwidth}}
\end{figure}

To take into account the finite bandwidth of the coupling modulations, we convolve the ideal square-wave train of pulses with a Gaussian filter.  In particular, we take the coupling to be given by
\begin{align}
 g(t)=\mathcal{F}^{-1}\left[g_\mrm{id}(\w)f(\w)\right],	\label{eqnFiltre1}
\end{align}
where
\begin{align}
 g_\mrm{id}(\w)=\mathcal F[g_\mrm{id}(t)],
 \hspace{5mm}
 f(\w)=\exp[-\w^2/2\s_\mrm f^2].	\label{eqnFiltre2}
\end{align}
In Eqs.~\eq{eqnFiltre1} and \eq{eqnFiltre2}, $\mathcal F^{(-1)}$ is the (inverse) Fourier transform, $g_\mrm{id}(t)$ is an ideal train of square-wave pulses having width $\tau^\prime$, period $2\tau$, and amplitude $g$.  The function $f(\w)$ is a Gaussian filter with standard deviation $\s_\mrm f$ which eliminates high-frequency components of $g_\mrm{id}(t)$. Evaluating Eq.~\eq{eqnFiltre1} gives
\begin{align}
 g(t)&=\sum_{j=0}^{n_p/2}g_\mrm{sq}(t-2j\tau),	\label{eqngdet}
\end{align}
where $g_\mrm{sq}(t)$ describes a single filtered square pulse centered around $t=0$,
\begin{align}
 g_\mrm{sq}(t)=\frac g2\left\{
  \mrm{erf}\left[
      \frac{\s_\mrm f}{\sqrt2}\left(t+\frac{\tau'}2\right)
    \right]
 -\mrm{erf}\left[
      \frac{\s_\mrm f}{\sqrt2}\left(t-\frac{\tau'}2\right)
    \right]
 \right\}.	\label{eqngsq}
\end{align}
In Eq.~\eq{eqngsq}, $\tau'$ may differ from the pulse interval $\tau$ [Fig.~\ref{figBandwidth}(a)]. Neglecting corrections that are suppressed exponentially with $\sigma_\mrm f\tau'$, the time required for the coupling to rise from $10\;\%$ to $90\;\%$ of its final value is, using Eq.~\eq{eqngsq}, 
\begin{align}
 t_\mrm r\simeq\frac{2\sqrt2\;\mrm{erf}^{-1}(4/5)}{\s_\mrm f}\simeq\frac{2.563103}{\s_\mrm f}.
 \label{eqntr}
\end{align}
When $\s_\mrm f < \w_\mrm q$, the spectral weight of $g(\w)$ is suppressed at $\w=2\w_\mrm q$ and the influence of the counter-rotating terms becomes negligible. When $\s_\mrm f>1/\tau^\prime \sim 1/\tau$, the pulse train $g(t)$ closely approximates a sequence of square-wave pulses.  Crucially, there is always the possibility for a separation between $\w_\mrm q$, $\s_\mrm f$, and $1/\tau$, allowing both conditions to be satisfied simultaneously:
\begin{align}
  \w_\mrm q>\s_\mrm f>\frac1\tau.	\label{eqnIneqFiltre}
\end{align}

To verify the analysis given above, we numerically evaluate the fidelity of a state transfer using Eq.~\eq{eqnDefF}, considering evolution under the toggling-frame Hamiltonian given in Eq.~\eq{eqnHtoggCR}, which fully accounts for counter-rotating terms. For a finite pulse rise time, $t_\mrm r\ne 0$, $g(t)$ will be non-zero even for $n(t)$ odd. To suppress the resulting unwanted excitations of the qubit and cavity, we then take $\tau'=\tau-t_\mrm r$ [$\tau'$ and $t_\mrm r$ are illustrated in Fig.~\ref{figBandwidth}(a)]. Consequently, the time-averaged coupling $\overline g=\int_0^{2\tau}dt\,g(t)/2\tau$ also decreases; the pulse interval that results in a complete state transfer is then obtained by numerically solving $\overline g n_p\tau=\pi/2$ for a given value of $g$ and $n_p$. The resulting state-transfer error is shown in Fig.~\ref{figBandwidth}(b) (blue dots) as a function of $\s_\mrm f/g$ for $n_p=100$, $gT_2^\ast=1/10$, and $\kp=0$. As $\s_\mrm f/g$ increases, the error decreases, approaching the value given by Eq.~\eq{eqnFid} for perfectly square modulation of $g(t)$ (dashed black line). Additional error due to a finite pulse rise time becomes negligible compared to the error already present for an ideal pulse (for this choice of parameters) for $\s_\mrm f\gtrsim500 g$. Even when the bandwidth is too narrow for saturation to have occurred, $\s_\mrm f<500 g$, Fig.~\ref{figBandwidth}(b) shows that the error due to inhomogeneous broadening can be suppressed substantially (without SQUADD, error would be of order 1 for $gT_2^\ast=1/10$). In the simulations, we have taken $\w_\mrm q=2000 g$. This choice allows us to filter out the effect of the counter-rotating terms over the entire parameter range of the simulation, guaranteeing that $\s_\mrm f<\w_\mrm q$ is satisfied. 

To give a concrete example, taking $g/2\pi=1$~MHz and $\s_\mrm f=100g$, the parameters used in the simulation presented in Fig.~\ref{figBandwidth}(b) correspond to $\s_\mrm f / 2\pi = 100$~MHz, $\w_\mrm q/2\pi = 2$~GHz, and $T_2^\ast\simeq16$~ns. Numerically solving $\overline g n_p\tau=\pi/2$ then gives $\overline g/2\pi\simeq0.27$~MHz and $\tau\simeq 9$~ns for $n_p=100$. Even for this narrow bandwidth (which leads to $t_\mrm r\simeq4$~ns), our simulation yields a relatively small error, $1-F\simeq 10^{-3}$.

\subsection{Finite $\pi$-pulse duration	\label{secImpulsionsFinies}}

In typical experimental settings, the finite duration of $\pi$-pulses poses an important practical limitation to the achievable fidelity of quantum operations~\cite{bluhm2011dephasing,viola2003robust,khodjasteh2007performance,pasini2011high,biercuk2009optimized,du2009preserving}. Therefore, in this section, we numerically evaluate the state-transfer fidelity under SQUADD as a function of the $\pi$-pulse duration.

With the cavity and the qubit on resonance, and working in the rotating frame, we describe qubit $\pi$-pulses with finite duration using the Hamiltonian
\begin{align}
 H_\mrm c(t)=\frac{w(t)}2\s_x,	\label{eqnHc}
\end{align}
valid under the rotating-wave approximation for the qubit drive. In Eq.~\eq{eqnHc}, we have introduced the drive amplitude, $w(t)$. Within the rotating-wave approximation for the qubit-cavity coupling (which is justified for $\s_\mrm f<\w_\mrm q$, as explained above), substituting $U_\mrm c(t) = \exp[-i\int_0^t dt_1 w(t_1)\s_x/2]$ into Eq.~\eq{eqnDefHT} leads to the toggling-frame Hamiltonian
\begin{align}
 &H_\mrm T(t)=\frac\xi2\left[\cos\q(t)\s_z+\sin\q(t)\s_y\right]\notag\\
  &+g(t)\left[\frac{1+\cos\q(t)}2(a^\dag\s_-+a\s_+)+\frac{1-\cos\q(t)}2(a^\dag\s_++a\s_-)
  +i\sin\q(t)\frac{a^\dag-a}2\s_z\right],	\label{eqnHTrotation}
\end{align}
where we have introduced the qubit rotation angle
\begin{align}
 \q(t)=\int_0^tdt_1\,w(t_1).	\label{eqnAngleRotation}
\end{align}

We take $w(t)$ to describe a train of identical pulses. For numerical evaluation, we consider perfect rectangular pulses of amplitude $w_0$ and duration $t_\mrm p$ centered around $(m+\frac12)\tau$, $m\;\in\;\mathbb N$ [Fig.~\ref{figBandwidth}(a)]. For $w_0 t_\mrm p=\pi$, evolution under $H_\mathrm{T}(t)$ leads to $\pi$-pulses. We retrieve the toggling-frame Hamiltonian for instantaneous $\pi$-pulses [Eq.~\eq{eqnHtoggling}] when taking the limit $t_\mrm p\rightarrow 0$ in $H_\mrm T(t)$ given by Eq.~\eq{eqnHTrotation} while requiring $w_0 t_\mrm p=\pi$. In plots and numerical evaluations, we take $g(t)$ to be given by Eq.~\eq{eqngdet}, above. The strategies proposed here for error suppression with a finite pulse duration are, however, independent of the specific shape of $w(t)$ and $g(t)$. Importantly, the counter-rotating [$\propto a^\dag\s_++a\s_-$] and cavity-displacement [$\propto (a^\dag -a)\s_z$] terms in Eq.~\eq{eqnHTrotation} both vanish for times $t$ satisfying $\q(t)=2j\pi$ (with $j\,\in\,\mathbb N$), at which the state-transfer is complete. These unwanted terms are, however, finite for $\q(t)\neq2j\pi$ [i.e., during $\pi$-pulses and for $n(t)$ odd]. The error that results can be suppressed by approximately turning off the coupling during the qubit $\pi$-pulses [by choosing, e.g., $\tau'=\tau-t_\mrm r-t_\mrm p$ as shown in Fig.~\ref{figBandwidth}(a)], in addition to turning the coupling off for $n(t)$ odd. Any error arising from a finite pulse duration is then predominantly due to the term $\propto\xi\sin\q(t)\s_y$ in Eq.~\eq{eqnHTrotation}.

As in Section~\ref{secBandePassante}, we numerically evaluate the fidelity $F$ [Eq.~\eq{eqnDefF}] of a quantum state transfer generated by the toggling-frame Hamiltonian given in Eq.~\eq{eqnHTrotation}. Figure~\ref{figBandwidth}(c) displays the resulting error $1-F$ as a function of $t_\mrm p$ (blue dots) for $n_p=100$, $gT_2^\ast=1/10$, $\s_\mrm f/g=1000$, $\tau'=\tau-t_\mrm r-t_\mrm p$, and $\kappa=0$. To suppress the error below $1\%$, $g t_\mrm p/2\pi<10^{-4}$ is required for these parameters. For, e.g., $g/2\pi=1$~MHz, this level of error suppression would require $t_\mrm p<0.1$~ns. Realizing such short pulses may be challenging experimentally.

To explain this poor suppression of error in the limit of small $t_\mrm p$, we note that, for the pulse sequence described below Eq.~\eq{eqnAngleRotation} [top-left inset in Fig.~\ref{figBandwidth}(c)], $H_\mrm T(t)$ does not result in a symmetric cycle (in the sense described in Section~\ref{secAHT}). Indeed, for this sequence with period $T=2\tau$, $\q(T-t)=2\pi-\q(t)$, leading to $\sin\q(T-t)=-\sin\q(t)$ and thus to $H_\mrm T(T-t)\neq H_\mrm T(t)$ [see Eq.~\eq{eqnHTrotation}]. We thus expect the error to be finite at first order in average Hamiltonian theory \cite{mehring1976high}, $\Hb1\neq0$ [see the discussion following Eq.~\eqref{eqnHmoyen2}]. This sequence only becomes a symmetric cycle [i.e., $H_\mrm T(T-t)= H_\mrm T(t)$, leading to $\Hb1=0$] in the limit $t_\mrm p\rightarrow0$. 

A symmetric cycle (leading to $\Hb1=0$) may be obtained using the phase-alternated sequence described in Section~\ref{secPulses}, in which the qubit rotation direction [i.e. the sign of $w(t)$] alternates from one $\pi$-pulse to the next. However, for this sequence, $\int_0^{T}dt_1 \sin\q(t_1)\neq0$, leading to finite error at zeroth order ($\Hb0$) for any pulse with finite $t_\mrm p$ due to the term $\propto\sin\q(t)\s_y$ in Eq.~\eq{eqnHTrotation}. A sequence that is a symmetric cycle (leading to $\Hb1=0$) and for which $\int_0^{T}dt_1 \sin\q(t_1)=0$ (leading to vanishing error at zeroth order) is obtained when:
\begin{enumerate}
 \item $w(t)$ is periodic and odd, $w(t)=-w(-t)$;
 \item $w(t)$ describes identical $\pi$-pulses that come in pairs with common phase, corresponding to successive rotations about $+\hat x,\;+\hat x,\;-\hat x,\;-\hat x,\;...$, with no specific assumption about the pulse shape except that $w(t)>0\;\forall\;t\in\;[0,\tau]$. 
\end{enumerate}
Error due to finite pulse duration then arises entirely from terms of order $\Hb2$ or higher. A sequence that fulfills the above criteria is shown in the bottom-right inset in Fig.~\ref{figBandwidth}(c), in which we consider the specific example of square pulses. 

As expected from the above discussion, the phase-alternated sequence shown in the bottom-right inset in Fig.~\ref{figBandwidth}(c) leads to an error [red triangles in Fig.~\ref{figBandwidth}(c)] that is significantly reduced compared with the original fixed-phase sequence [blue dots in Fig~\ref{figBandwidth}(c)]. The error is also significantly more strongly suppressed in the limit of small $t_\mrm p$ for the phase-alternated sequence, relative to the fixed-phase sequence. For the same parameters as above, an error $<1\%$ is obtained for $g t_\mrm p/2\pi=10^{-3}$, corresponding to $t_\mrm p=1$~ns for $g/2\pi=1$~MHz. For $gt_\mrm p/2\pi<10^{-3}$, the error quickly decreases and plateaus near the value given by Eq.~\eq{eqnFid}. Finally, as explained in Section~\ref{secPulses}, phase-alternated sequences have the additional advantage of suppressing an accumulation of deterministic pulse errors.

\section{Qubit readout	\label{secReadoutMain}}
Recently, a longitudinal qubit-cavity interaction $[\propto g(a^\dag +a)\s_z]$ has been considered theoretically and shown to produce a quantum nondemolition readout that is faster than the usual dispersive readout~\cite{didier2015fast}.
Here, we show how this type of interaction can be engineered simply by applying the Carr-Purcell sequence on a qubit with a fixed coupling to the cavity, $g(t)=g\;\forall\;t$, leading to qubit readout.
To simplify the discussion, throughout this section we neglect contributions from inhomogeneous broadening [i.e., we take $\xi=0$ in Eq.~\eq{eqnHtoggling}].

If the Carr-Purcell sequence shown in Fig.~\ref{figSequence} is applied with a fixed coupling, $g(t)=g\;\forall\;t$, the counter-rotating term in Eq.~\eqref{eqnHtoggling} contributes. Although this is harmful to state transfer, this term can also generate otherwise useful quantum operations. Indeed, the evolution operator from leading-order average Hamiltonian theory is then
\begin{align}
 U_\mrm R(t_f)=\eul{-ig(a^\dag+a)\s_x t_f/2}= D(-i\s_x gt_f/2),	\label{eqnULecture}
\end{align}
where $D(\al)$ is the displacement operator producing the coherent state $\ket{\al}_\mrm c\equiv D(\al)\ket 0_\mrm c$~\cite{scully1997quantum}.
The interaction appearing in Eq.~\eq{eqnULecture} is longitudinal with respect to $\s_x$ eigenstates, $\ket{\pm}_\mrm q$. Applying $U_\mrm R(t_f)$ on $\ket \pm_\mrm q\ket0_\mrm c$ then gives
\begin{align}
 U_\mrm R(t_f)\ket\pm_\mrm q\ket 0_\mrm c=\ket\pm_\mrm q{\ket{\pm\al}_\mrm c},	\label{eqnDisplacement}
\end{align}
with $\al\equiv -igt_f/2$.
Thus, in combination with a qubit rotation conditioned on the cavity state~\cite{leghtas2013deterministic}, $U_\mrm R(t_f)$ can be used to map a qubit state to a superposition of cavity coherent states; a Schr\"odinger's cat state~\cite{vlastakis2013deterministically,wang2015decoherence}. Alternatively, the states $\ket{\pm\al}_\mrm c$ can be resolved by homodyne detection of the signal leaking from the cavity, enabling quantum nondemolition readout of the qubit in the basis $\{\ket\pm_\mrm q\}$. 

In the rest of this section, we  investigate the limitations of the qubit readout resulting from this combination of Hamiltonian engineering and homodyne detection.  For averaged measurements, the appropriate figure of merit is the signal-to-noise ratio of an associated estimator for the qubit expectation value.  For single-shot measurements, the appropriate figure of merit is the single-shot measurement fidelity.  We will find that these two measures can indicate a high-quality readout for this protocol even in the weak-coupling regime ($g<\kp$).

\subsection{Signal-to-noise ratio	\label{secSNR}}

When a qubit is successively prepared and measured $m\gg 1$ times to estimate an expectation value, the measurement statistics describing the mean of many independent repeated measurements (a so-called `soft average' \cite{d2014soft,ryan2015tomography}) become Gaussian due to the central limit theorem. The performance of the readout is then well-characterized by the signal-to-noise ratio (SNR). For a qubit being measured through a cavity, as considered here, the SNR compares the first two moments of the measurement operator
\begin{align}
 M=i\sqrt \kp\int_0^{t_f}dt[a_\mrm{out}^\dag(t)-a_\mrm{out}(t)],	\label{eqnDefM}
\end{align}
which gives the integrated homodyne-detection signal for a measurement time $t_f$, with $a_\mrm{out}(t)$ the output field leaking from the cavity with damping rate $\kp$. These first two moments of $M$ are quantified by the measurement signal $X$ and noise $\Xi$,
\begin{align}
 X&=|\mean M_+-\mean M_-|,	\label{eqnDefSignal}\\
 \Xi&=(\D M_+^2+\D M_-^2)^{1/2},	\label{eqnDefNoise}\\
 \D M_\pm^2&=\mean{M^2}_\pm-\mean{M}_\pm^2,	\label{eqnDefVar}
\end{align}
where $\mean{O}_\pm\equiv\tr[O(t_f) \rho_\pm(0)]$ and $\rho_\pm(0)\equiv\proj\pm_\mrm q\otimes\proj0_\mrm c$.
The signal-to-noise ratio is then simply
\begin{equation}
 \SNR=X/\Xi.
\end{equation}
In this section, we evaluate $X$ and $\Xi$ for the readout scheme described below Eq.~\eq{eqnDisplacement}, accounting for the first two nonvanishing orders in the Magnus expansion for the time-periodic Liouvillian $\liouv$ [defined explicitly in Eq. \eqref{eqnMaster}, below]: $\overline\liouv = \Lb0+\Lb2$. We will show that while $\Lb{0}$ generates the required conditional coherent-state displacement, $\Lb{2}$ results in qubit switching at a rate $\G\simeq g^2\tau^2\kp/24$ in the basis $\{\ket\pm_\mrm q\}$. This qubit switching acts as a source of telegraph noise in the Langevin equation for the cavity field $a(t)$~\cite{gardiner2000quantum}. We will evaluate the SNR including noise fom qubit switching and show that for a sufficiently short pulse interval, $\kp\tau<1$, the readout considered here can result in a large SNR $(\mathrm{SNR}> 1)$ even in the weak-coupling regime ($g<\kp$). 

To evaluate the SNR for a given measurement scheme, it is useful to relate $M$ to the input field $a_\mrm{in}(t)$ and to the field $a(t)$ inside the cavity. This relation is given by the input-output formula~\cite{gardiner2000quantum}
\begin{align}
 a_\mrm{out}(t)=a_\mrm{in}(t)+\sqrt \kp \,a(t).	\label{eqnEntreeSortie}
\end{align}
Assuming that the input is vacuum, substitution of Eq.~\eq{eqnEntreeSortie} into Eq.~\eq{eqnDefM} gives
\begin{align}
 \mean{M}_\pm&=i\kp\int_0^{t_f}dt[\mean{a^\dag(t)}_\pm-\mean{a(t)}_\pm],	\label{eqnM}\\
 \mean{M^2}_\pm&=\kp t_f
 +2\kp^2\int_0^{t_f}dt_1\int_0^{t_f-t_1}\!\!\!dt_2\left[
  \mean{a^\dag(t_1+t_2)a(t_1)}_\pm
  -\mean{a(t_1+t_2)a(t_1)}_\pm+\hc\right].	\label{eqnM2}
\end{align}
Eqs.~\eq{eqnM} and \eq{eqnM2} relate $\mean{M}_\pm$ and $\mean{M^2}_\pm$---and thus the SNR---to simple expectation values and autocorrelation functions of the cavity field $a(t)$. Employing standard formulas~\cite{gardiner2000quantum}, these expectation values and autocorrelation functions are easily calculated knowing the time-evolution superoperator $V(t,t_0)$, generating evolution for the qubit-cavity density matrix.
We find $V(t,t_0)$ by solving the (time-inhomogeneous) master equation
\begin{align}
 \dot{V}(t,t_0)=\liouv(t)V(t,t_0).	\label{eqnMaster}
\end{align}
In Eq.~\eq{eqnMaster}, we have introduced the Liouvillian $\liouv(t)$ describing cavity damping at rate $\kp$ and unitary evolution under the qubit-cavity toggling-frame Hamiltonian, $H_\mrm{T}(t)$,
\begin{align}
 \liouv(t)\cdot&=-i[H_\mrm T(t),\cdot]+\kp\mathcal D[a]\cdot	\label{eqnLiouv},\\
 \mathcal{D}[a]\cdot&=a\cdot a^\dag-\textstyle{\frac12}\left(a^\dag a \cdot+\cdot a^\dag a\right),
\end{align}
where the centerdot (``$\cdot$'') represents an arbitrary operator upon which the relevant superoperator is applied. 
In Eq.~\eq{eqnLiouv}, $H_\mrm T(t)$ is given by Eq.~\eq{eqnHtoggling}, taking $\xi=0$ and $g(t)=g\,\forall\,t$.

To evaluate $V(t,t_0)$ analytically, we assume that ${t-t_0=2n_p\tau}$. We then use the Magnus expansion
\begin{align}
 V(t,t_0)&=\exp\left[(t-t_0)\sum_{k=0}^\infty\Lb{k}\right].	\label{eqnMagnusV}
\end{align}
As in average Hamiltonian theory, the terms $\Lb{k}$ are time-independent because $\liouv(t)$ is periodic, $\liouv(t+2\tau)=\liouv(t)\,\forall\, t$. 
The first few terms of the expansion in $\Lb k$ are obtained by replacing $\Hb k\rightarrow i \Lb k$ and $H_\mrm T(t)\rightarrow i \liouv(t)$ in Eqs.~\eq{eqnHmoyen0} to~\eq{eqnHmoyen2}~\cite{blanes2009magnus}.

To gain insight into the problem, we evaluate the SNR to leading order in the Magnus expansion, Eq.~\eq{eqnMagnusV}.
In this first approach, we neglect any qubit decay that may arise from higher-order terms in the expansion. We then find
\begin{align}
 V(t,t_0)&=\exp[(t-t_0)\Lb{0}],	\label{eqnVL0}\\	
 \Lb{0}\cdot&=-i[\Hb{0},\cdot]+\kp\mathcal D[a]\cdot,	\label{eqnL0}\\
 \Hb{0}&=\frac g2(a+a^\dag)\s_x.	\label{eqnH0lecture}
\end{align}
According to Eqs.~\eq{eqnVL0} and \eq{eqnL0}, the qubit forever remains in its initial state $\proj\pm_\mrm q$. 
For $\kp t_f\gg1$, the cavity correspondingly settles in the coherent state $\proj{\al}_\mrm c=\proj{\mp i g/\kp}_\mrm c$.
Since the cavity field leaks from the output port at a rate $\kp/2$, this steady state leads to $X\propto \kp t_f\times g/\kp\propto g t_f$.
In addition, noise in the output field then entirely comes from shot noise: $\D M^2_\pm=\kp t_f$, giving~\cite{didier2015fast}
\begin{align}
 X&=4gt_f,	\hspace{3mm}	\Xi=\sqrt{2\kp t_f}
 \hspace{2.5mm}\Rightarrow\hspace{2.5mm}\mrm{SNR}\propto\sqrt{t_f}.	\label{eqnSNRideal}
\end{align}
Therefore, in this ideal scenario, signal always accumulates faster than noise, making it possible to achieve arbitrarily large SNR simply by increasing $t_f$. 

In practice, qubit relaxation leads to a saturation of the signal and to an enhancement of the noise, thus limiting the achievable SNR. Qubit relaxation can be intrinsic, coming from coupling of the qubit to a decay channel independent of the cavity. Higher-order corrections to the leading-order Magnus expansion taken here also lead to qubit decay via the cavity. This can be seen by means of a short-time expansion of $\mean{\s_x(t)}_\pm$. Indeed, the term of order $\ordre(t)$ in this short-time expansion gives decay at a rate analogous to that of Purcell decay:
\begin{align}
 \G \equiv\left|\frac d{dt}\mean{\s_x(t)}_\pm \right|_{t\rightarrow0}\simeq\left|\tr\{[\Lb{2}\,\!^\dag\s_x]\rho_\pm(0)\}\right|=\frac{g^2\tau^2}{24}\kp.
 \label{eqnGamma}
\end{align}
The term of order $\ordre(t)$ in the above expansion of $\mean{\s_x(t)}_\pm$ dominates over the correction term of order $\ordre(t^2)$ when
$\kp t< 256/[3(\kp\tau)^2]$.

\begin{figure*}
 \begin{center}
 \includegraphics[width=0.48\textwidth]{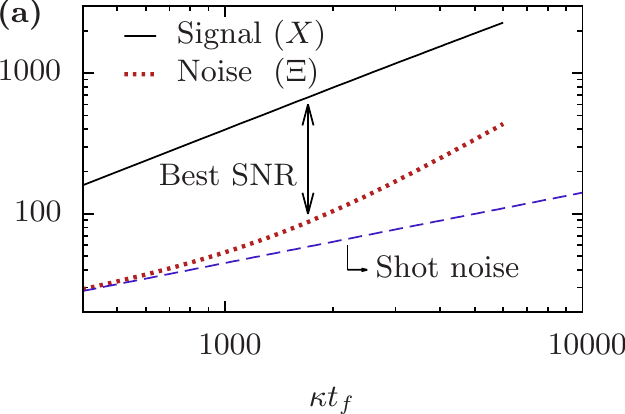}
 \raisebox{1.5mm}{\includegraphics[width=0.48\textwidth]{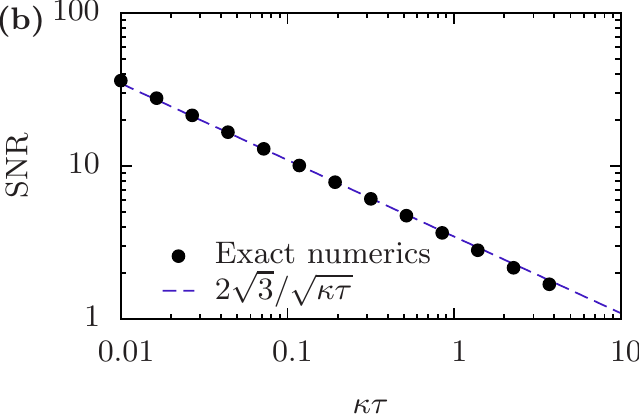}}
 \end{center}
 \caption{Signal-to-noise ratio (SNR) for the proposed readout with $g/\kp=1/10$. 
 (a) Dynamics of signal and noise accumulation for measurement time $t_f$. Solid black line: measurement signal $X$, Eq.~\eq{eqnDefSignal}. Red dotted line: measurement noise $\Xi$, Eq.~\eq{eqnDefNoise}. Dashed blue line: $\Xi$ for shot noise only, Eq.~\eq{eqnSNRideal}. The double arrow indicates the measurement time that optimizes the ratio $\SNR=X/\Xi$. $X$ and $\Xi$ are evaluated for $\kp\tau=0.2$. 
 (b)~Maximal SNR as a function of $\kp\tau$.
 In (a) and (b), $X$ and $\Xi$ are evaluated using a numerical solution of the master equation, Eq.~\eq{eqnMaster}. For $g/\kp=1/10$, a cavity Hilbert space of dimension 3 is sufficient for accurate numerical evaluation of $X$ and $\Xi$.
 \label{figSNR}}
\end{figure*}

To take qubit relaxation into account in the calculation of the SNR, we employ the Langevin equation for the cavity field $a(t)$, considering the average Hamiltonian $\Hb{0}$ in Eq.~\eq{eqnH0lecture}. This gives
\begin{align}
 \dot a(t)+\frac\kp2a(t)=-i\frac g2\s_x(t)-\sqrt\kp a_\mrm{in}(t).	\label{eqnLangevin}
\end{align}
Eq.~\eq{eqnLangevin} has the form of the equation of motion of a Brownian particle with mass $m$, momentum $p$, and friction coefficient $\g$: $\dot p+(\g/m)p=\eta(t)$~\cite{gardiner2000quantum}. In Eq.~\eq{eqnLangevin}, the fluctuating force $\eta(t)$ comes from a combination of shot noise from the input field $a_\mrm{in}(t)$ and telegraph noise from the qubit through the Heisenberg-picture operator $\s_x(t)$. We assume that the qubit-cavity coupling is turned on at time $t=0$ and that the cavity interacts with its environment starting in the distant past, at $t\rightarrow-\infty$. The solution to Eq.~\eq{eqnLangevin} is then
\begin{align}
 a(t)&=-i\frac g2\int_0^tdt'\eul{-\kp(t-t')/2}\s_x(t')	
  -\sqrt\kp\int_{-\infty}^tdt'\eul{-\kp(t-t')/2}a_\mrm{in}(t').	\label{eqnSola}
\end{align}
For a qubit undergoing simultaneous excitation and relaxation at equal rates $\G/2$ in the eigenbasis of $\s_x$, we have
\begin{align}
 &\mean{\s_x(t)}_\pm=\pm\exp(-\G t),	\label{eqnsx}\\
 &\mean{\s_x(t)\s_x(t')}_\pm=\exp(-\G|t-t'|).	\label{eqnsxsx}
\end{align}
Substituting Eqs.~\eq{eqnSola} to \eq{eqnsxsx} into Eqs.~\eq{eqnM} and \eq{eqnM2}, we evaluate the signal and noise using Eq.~\eq{eqnDefSignal} and Eq.~\eq{eqnDefNoise}. We find
\begin{align}
 X=\frac{2g\kp}{\kp/2-\G}\left(\frac{1-\eul{-\G t_f}}\G-\frac{1-\eul{-\kp t_f/2}}{\kp/2}\right),	\label{eqnX1}
\end{align}
\begin{align}
 \Xi&=\sqrt{2\kp t_f+\frac{4g^2\kp^2}{(\kp^2/4-\G^2)\G^2}f(t_f)-\frac{X^2}2},	\label{eqnXi1}
\end{align}
where we have introduced
\begin{align}
 &f(t)=\G t-\frac\G{\G+\kp/2}\left(1+\frac{2\G}\kp\right)\left[1-\eul{-(\G+\kp/2)t}\right]
      -\left(1-\frac{2\G}\kp\eul{-\kp t/2}\right)\left(1-\eul{-\G t}\right)\notag\\
 &\;\;\;+\frac{2\G}\kp\left(1-\eul{-\kp t/2}\right)\left[\eul{-\G t}+\frac\G\kp\left(1-\eul{-\kp t/2}\right)\right]
      -\frac{4\G^2}{\kp^2}\left[\G t-\frac\G\kp\left(3-4\eul{-\kp t/2}+\eul{-\kp t}\right)\right].	\label{eqnf}
\end{align}
To simplify the above expressions, we expand $X$ and $f(t_f)$ to leading order in $\G t_f$ around $\G t_f=0$. We also assume that the cavity has reached its steady state; we thus have $\G t_f\ll1\ll\kp t_f$. Therefore, in Eqs.~\eq{eqnX1} and \eq{eqnf}, we drop corrections that are exponentially small for $\kp t_f\gg1$.
In Eq.~\eq{eqnX1}, we also drop terms of order $\mathcal O(\G/\kp)$ or higher, which do not change the dependence of $X$ and $\Xi$ on $t_f$. However, in Eq.~\eq{eqnf}, we keep the terms of order $\mathcal O(\G/\kp)$, since they grow faster than linearly with $t_f$, but drop corrections of order $\mathcal O(\G^2/\kp^2)$ or higher. We then find
\begin{align}
 X\simeq4gt_f,\hspace{5mm}
 \Xi\simeq\sqrt{2\kp t_f+\frac{16}3g^2\G t_f^3}.	\label{eqnXXi}
\end{align}
Equation~\eq{eqnXXi} shows that $\Xi^2$ contains two terms: one from photon shot noise, $\propto \kp t_f$, and an additional contribution from qubit switching, $\propto g^2\G t_f^3$.
Therefore, including qubit switching, the noise grows faster than the signal ($\propto t_f$) for sufficiently large $t_f$. 
This is visible in Fig.~\ref{figSNR}(a), in which we plot $X(t_f)$ and $\Xi(t_f)$ resulting from an exact numerical solution of the master equation given by Eq.~\eq{eqnMaster}. In Fig.~\ref{figSNR}(a), $X(t_f)$ and $\Xi(t_f)$ are represented by the solid black line and the dotted red line, respectively. Using the dashed blue line, we also plot $\Xi(t_f)=\sqrt{2\kp t_f}$, expected for pure photon shot noise, Eq.~\eq{eqnSNRideal}. Clearly, excess noise due to qubit decay determines the optimal measurement time $t_\mrm{opt.}$ that maximizes the SNR [shown by the double arrow in Fig.~\ref{figSNR}(a)]. We evaluate $t_\mrm{opt.}$ analytically by maximizing $\mrm{SNR}=X/\Xi$, with $X$ and $\Xi$ given by Eq.~\eq{eqnXXi}. We find
\begin{align}
 \G t_\mrm{opt.}&\simeq\frac12\sqrt{\frac32}\frac{\sqrt{\kp\G}}{g},	\label{eqntopt}\\
 \mrm{SNR}&\simeq\left(\frac{6g^2}{\kp\G}\right)^{1/4}.
 \hspace{1.5cm}(t_f=t_\mrm{opt.})\label{eqnSSBopt}
\end{align}
Equation~\eq{eqnSSBopt} provides a simple relationship between the maximal SNR and the cooperativity $C\equiv g^2/\kp\G$, $\mrm{SNR}\simeq(6C)^{1/4}$.

Equation \eq{eqnSSBopt} gives the maximal SNR when the qubit undergoes switching in the eigenbasis of $\s_x$.
As seen above, this can be due to the subleading term $\Lb{2}$ in the Magnus expansion, Eq.~\eq{eqnMagnusV}, which leads the qubit to decay at the rate given in Eq.~\eq{eqnGamma}.
When this mechanism is the dominant source of decay in the eigenbasis of $\s_x$, we substitute Eq.~\eq{eqnGamma} into Eq.~\eq{eqnSSBopt} to find the corresponding optimal SNR,
\begin{align}
 \mrm{SNR}\simeq\frac{2\sqrt3}{\sqrt{\kp\tau}},	\hspace{1.5cm}(t_f=t_\mrm{opt.})	\label{eqnSNRopt}
\end{align}
valid for $\G t_f\ll1\ll\kp t_f\ll256/[3(\kp\tau)^2]$. The last inequality arises from the short-time expansion performed in Eq.~\eq{eqnGamma}.
Equation~\eq{eqnSNRopt} implies that $\mrm{SNR}>1$ is achievable even in the weak-coupling regime, $g<\kp$.
This result is shown in Fig.~\ref{figSNR}(b), in which we plot the maximal SNR obtained from an exact numerical solution of Eq.~\eq{eqnMaster} as the black dots for $g/\kp=1/10$. This numerical result is in good agreement with the optimal SNR given in Eq.~\eq{eqnSNRopt}, displayed as the dashed blue line.

We now discuss conditions under which the Magnus expansion used here [Eq.~\eq{eqnMagnusV}] converges. The Magnus expansion converges when $\int_0^{2\tau}dt\|\liouv(t)\|<\pi$~\cite{blanes2009magnus}. For $g<\kp$, the steady-state cavity population is small: $\mean{a^\dag a(t)}= (g/\kp)^2<1$ for $\kp t_f\gg1$. In this situation, we can represent the operators $a$ and $a^\dag$ by truncated matrices of small dimension, making $\|a^{(\dag)}\|\sim1$. This implies that $\|\liouv(t)\|\sim \kp$, and we conclude that the Magnus expansion converges for $\kp\tau\lesssim\pi/2$ under the assumption that $g<\kp$. This statement is consistent with Fig.~\ref{figSNR}(b), which shows excellent agreement between the exact numerical solution (black dots) and Eq.~\eq{eqnSNRopt} (dashed blue line) for $\kp\tau<\pi/2$.

\subsection{Single-shot fidelity\label{secSSFidelity}}

In contrast to the case of many repeated measurements (described above), for a single-shot readout, the measurement statistics are typically non-Gaussian. Indeed, while the conditional probability distribution describing the integrated signal $\mean M_\pm$ would simply describe a displaced Gaussian in the absence of switching, random switching events (e.g. qubit decay due to the mechanism described above) lead to significant bimodality \cite{d2014optimal}. To characterize readout, the full probability distribution of the measurement outcomes is then needed; the first and second moments [characterized by Eqs.~\eq{eqnDefSignal} and \eq{eqnDefNoise}] are typically not sufficent. A good measure of quality that accounts for the full probability distribution is the single-shot fidelity. To evaluate the fidelity, we use a readout model that takes into account qubit switching at symmetric rates $\G/2$, where $\G$ is given by Eq.~\eq{eqnGamma}~\cite{d2014optimal,gambetta2007protocols}. In the same regime as above ($g\ll\kp$ and $\kp\tau\ll1$), this leads to a single-shot fidelity that converges asymptotically to
\begin{align}
 F_1=1-\frac{(\kp\tau)^2}{192}\left[\log\frac{96}{(\kp\tau)^2}+\mathcal O\left(|\log\kp\tau|^{-1/2}\right)\right]	\label{eqnFidReadout}
\end{align}
as $\kp\tau\rightarrow0$. For $\kp\tau=0.1$, this yields  a single-shot fidelity of $99.95\,\%$, showing that the error due to the first correction term in the Magnus expansion is rapidly suppressed in the limit of short pulse intervals. 

Equation~\eq{eqnFidReadout} also shows that the readout proposed here can have a high single-shot fidelity even in the weak-coupling regime ($g<\kappa$). This readout may then be useful in several novel experimental settings where it is challenging to achieve strong coupling.
For example, a spin qubit in a carbon nanotube has recently been successfully coupled to a microwave resonator, but the coupling achieved is marginal, $g/\kp\sim1$~\cite{viennot2015coherent}. 
Alternative setups for semiconductor spin qubits in quantum dots or at single donor impurities coupled to microwave cavities have predicted couplings $g/2\pi\lesssim 1$ MHz~\cite{hu2012strong,tosi2014circuit,tosi2015silicon}, typically smaller than the damping rate $\kp/2\pi=2$-$10$~MHz~\cite{frey2012dipole,basset2014evaluating}.

\section{Conclusion}

In summary, we have introduced and assessed protocols for two quantum operations relevant to cavity QED: (i) quantum state transfer between a qubit and a cavity, and (ii) qubit readout through the cavity output field. 

For quantum state transfer, the protocol presented here (SQUADD) can lead to a high fidelity even in the limit of strong dephasing due to inhomogeneous broadening. This result holds also when storing the logical qubit in a collective mode of a large ensemble of $N$ physical qubits. To evaluate the state-transfer fidelity for ensembles of physical qubits, we have shown that the dynamics of the state transfer under SQUADD is well approximated in a closed subspace formed by only four collective modes in the limit $N\gg1$. For quantum state transfer between a cavity and a single physical qubit, we have considered error arising from a finite off/on ratio in the tunable qubit-cavity coupling, deterministic $\pi$-pulse errors, finite bandwidth of the coupling pulses, and finite duration of the qubit $\pi$-pulses. We have also considered phase-alternated versions of SQUADD with reduced error from finite pulse duration and deterministic over (under)-rotations. 

We have shown that applying the Carr-Purcell sequence on a qubit with constant coupling to a cavity leads to a longitudinal interaction that can be used to produce a fast qubit readout (compared with the dispersive readout that is commonly used in cavity QED). This readout can have a large signal-to-noise and a high single-shot fidelity even in the weak-coupling regime. The above results for quantum state transfer and qubit readout are especially relevant to spin qubits~\cite{viennot2015coherent,bluhm2011dephasing,hu2012strong,kubo2012storage}, for which coupling to the cavity is typically weak compared with inhomogeneous broadening and/or cavity damping.

Since SQUADD builds on the well-known Carr-Purcell sequence, it may be easily incorporated into near-term experiments. In the future, applying more complex dynamical decoupling sequences (e.g., Uhrig~\cite{uhrig2007keeping} or concatenated~\cite{khodjasteh2005fault} dynamical decoupling) to a qubit coupled to a cavity may be a promising avenue for both quantum state transfer and qubit readout. Indeed, these alternative protocols may allow for a better suppression of error due to qubit dephasing and cavity-mediated qubit switching.

Moving forward, the ideas presented here could lead to applications well beyond state transfer and readout. For example, going to second order in average Hamiltonian theory yields terms $\propto i g^2\tau[a^2-(a^\dag)^2]$, which could be used to generate cavity squeezing (see Section~\ref{secOffOn}). Such squeezing may be useful, e.g., to further improve qubit readout~\cite{didier2015fast}, or to realize a high-fidelity two-qubit gate~\cite{royer2016fast,puri2016high}. In addition, by monitoring the coherence of a state that is periodically swapped between a qubit and a bosonic mode, it may be possible to characterize noise processes affecting a harmonic system (e.g., a cavity or a magnon mode~\cite{tabuchi2015coherent}). This may allow for the application of noise spectroscopy methods~\cite{bylander2011noise} to harmonic systems in situations where dynamical decoupling through parity kicks~\cite{vitali1999using} may be challenging.

\section*{Acknowledgments} 
We thank Nicolas Didier, Benjamin d'Anjou, Hugo Ribeiro, Michel Pioro-Ladri\`ere, and Dany Lachance-Quirion for useful discussions. This research was undertaken thanks in part to funding from the Canada First Research Excellence Fund. This work was supported by NSERC, CIFAR, and the W. C. Sumner Foundation.



\section*{References}

\bibliographystyle{iopart-num}
\bibliography{article}

\end{document}